\documentclass[journal]{IEEEtran}

\usepackage{balance}
\usepackage{amssymb}
\usepackage[normalem]{ulem}
\usepackage{multirow}
\usepackage{subfigure}
\usepackage{cite}
\usepackage[dvips]{graphicx}
\usepackage{amsmath}
\usepackage{url}
\usepackage{psfrag}
\usepackage{amsbsy}
\usepackage{rsfs}
\usepackage{verbatim}
\usepackage{rotating}
\usepackage{booktabs}
\usepackage{xcolor}
\usepackage{flushend}

\renewcommand{\j}{\mathrm{j}}
\newcommand{\R}{\mathrm{R}}
\newcommand{\T}{\mathrm{T}}
\newcommand{\tot}{\mathrm{tot}}

\newcommand{\threedB}{\mathrm{3dB}}

\newcommand{\Ang}[1]{${#1}^{\circ}$}

\newcommand{\Vect}[1]{\boldsymbol{#1}}

\newif\ifMarking
\ifMarking
    \newcommand{\revision}[4]{\colorbox{#1}{#2}{\color{gray}{#3}}{\color{blue}{#4}}}
\else
    \newcommand{\revision}[4]{#4}
\fi

\begin{document}

\title{Synthesized-Isotropic Narrowband Channel Parameter Extraction from Angle-Resolved Wideband Channel Measurements}

\author{Minseok~Kim,~\IEEEmembership{Senior Member,~IEEE,}
        ~Masato~Yomoda
\thanks{M. Kim and M. Yomoda are with the Graduate School of Science and Technology, Niigata University, Niigata 950-2181, Japan (e-mail: mskim@eng.niigata-u.ac.jp).}
\thanks{This work was partly supported by the Commissioned Research through the
National Institute of Information and Communications Technology (NICT)
(\#JPJ012368C02701), and the Ministry of Internal Affairs and Communications (MIC)/FORWARD (\#JPMI240410003), Japan.}
}

\maketitle

\begin{abstract}
Angle-resolved channel sounding using antenna arrays or mechanically steered high-gain antennas is widely employed at millimeter-wave and terahertz bands. To extract antenna-independent large-scale channel parameters such as path loss, delay spread, and angular spread, the radiation-pattern effects embedded in the measured responses must be properly compensated. This paper revisits the technical challenges of path loss/path gain calculation from angle-resolved wideband measurements, with emphasis on angular-domain power integration where the scan beams are inherently non-orthogonal and simple power summation leads to biased isotropic-equivalent power estimates. We first formulate the synthesized-isotropic narrowband power in a unified matrix form and introduce a beam-accumulation correction factor, including an offset-averaged variant to mitigate scalloping due to off-grid angles. The proposed framework is validated through simulations using channel models and 154~GHz corridor measurements.
\end{abstract}


\section{Introduction}
Path loss (PL), defined as the reciprocal of path gain (PG), is a fundamental large-scale parameter for radio system development \cite{Rappaport,Erceg}. In principle, antenna-independent PL should be determined only by the propagation channel and therefore corresponds to an isotropic-antenna assumption \cite{ITU_R_P_341,ITU_R_P_1407}. In practical millimeter-wave and terahertz measurements, however, omnidirectional sounding is often infeasible because of the limited link budget. Instead, angle scanning with high-gain antennas or antenna arrays is commonly used to resolve the angular structure of multipath propagation. Accordingly, double-directional (D-D) channel sounding employs mechanical steering or array-based beamforming at both the transmitter (Tx) and receiver (Rx) \cite{Steinbauer,KIM_11GIndoor,Zwick,Kim_IEEEAccess2021_FullAzSweep}. A main challenge in such measurements is how to derive isotropic-equivalent channel parameters, such as path loss/path gain and temporal/angular power dispersions, from the measured angle-resolved wideband data.

Conceptually, isotropic-equivalent power recovery requires integration of the measured power over delay and angle. In the delay domain, this is usually well behaved after standard back-to-back calibration \cite{Thoma,Doeker_OJAP2024_Calibration}, and summation over delay bins provides a consistent estimate of the wideband received power. The angular domain is more difficult. Practical scan beams generated by arrays or horn antennas are not orthogonal, and adjacent pointings often overlap substantially. As a result, the integrated angular power depends strongly on the angular scanning interval (ASI). When the ASI is too coarse, energy arriving between beam centers is attenuated by scalloping loss, leading to underestimation of the isotropic-equivalent power. When the ASI is too fine, neighboring pointings overlap heavily, and naive power summation counts correlated contributions from the same path multiple times, leading to overestimation.

To address these issues, parametric high-resolution estimators such as Space-Alternating Generalized Expectation-maximization (SAGE) \cite{Fleury_SAGE}, Richter's Maximum Likelihood (RIMAX) \cite{Richter_RIMAX}, and CLEAN algorithm \cite{Kim_CLEAN,Kim_Springer} incorporate antenna responses to estimate multipath parameters, including Angle of Arrival (AoA), Angle of Departure (AoD), delay, and complex weights, from which an antenna-independent path gain can be obtained by summing the estimated path powers. In practice, however, the multipath components captured by a high-resolution estimator depend on the adopted high-resolution estimator and its underlying modeling assumptions. While some approaches primarily emphasize a limited number of dominant specular components, others can also account for diffuse or dense multipath components (DMC) \cite{Kaske}. Accordingly, the extent to which the total received power is captured may vary depending on the estimation framework and channel conditions, which in some cases motivates supplementary omnidirectional measurements for PL characterization \cite{KIM_11GIndoor}. By contrast, non-parametric approaches (e.g., beamforming-based angular power spectra) are computationally robust but remain biased due to non-orthogonal beams and sidelobe-induced angular leakage, which distorts angular integration and the resulting isotropic-equivalent power.

Several studies have addressed isotropic-equivalent power and PL estimation from directional measurements. A practical synthesis of omnidirectional received power and PL from directional scanning measurements has been proposed by
combining measurements over multiple pointing angles \cite{Sun_Globcom2015_OmniSynth}. The fundamental difficulty of estimating omnidirectional PL from directional sounding, where the same propagation paths may appear in multiple pointings and thus complicate direct power aggregation, has also been discussed
\cite{Haneda_EuCAP2016_OmniPL}. A linear combining framework that accounts for antenna gain patterns and nonuniform angular sampling has been studied for isotropic PL estimation \cite{Daoud}. High-resolution multipath angle estimation has been developed to improve parameter extraction from power--angle--delay profiles without increasing measurement complexity \cite{Xu}. In addition, the virtual antenna array (VAA) approach has been investigated for omnidirectional PL measurement using directional antennas \cite{Li_Fan_TVT2023}. 

Compared with these approaches, the proposed method is intended for a different purpose. Rather than relying on explicit high-resolution multipath parameter extraction or more elaborate inversion procedures, it provides a simple and implementation-friendly correction framework for synthesized-isotropic path loss/path gain estimation directly from discretely sampled angle-resolved measurements. Its main practical advantage is that beam-overlap and scalloping effects are compensated through a beam-accumulation correction factor that can be computed directly from the scan grid and antenna pattern, making the method attractive for practical directional-scanning campaigns where robustness and ease of use are important.

\begin{figure*}[t]
\centering
\subfigure[Isotropic measurement.\label{fig:omni_meas}]{\includegraphics[width=0.34\linewidth]{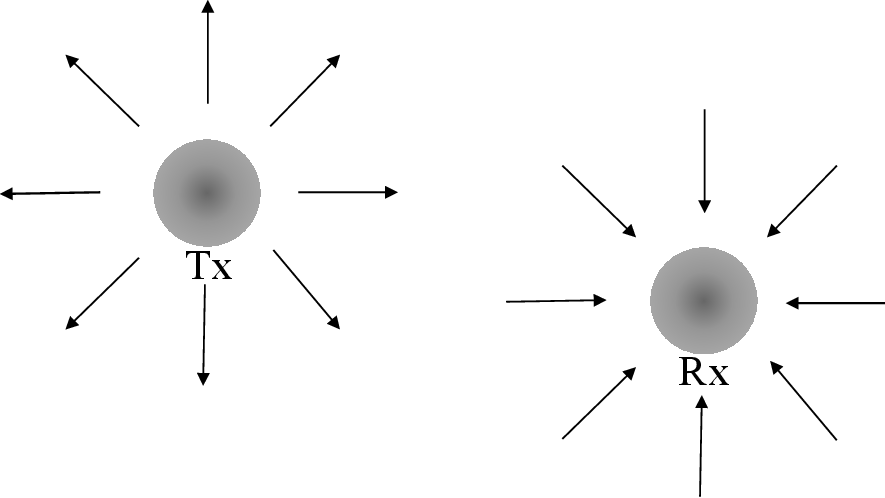}} 
\qquad
\subfigure[Double-directional angle-scanning.\label{fig:directional_scan_spherical}]{\includegraphics[width=0.6\linewidth]{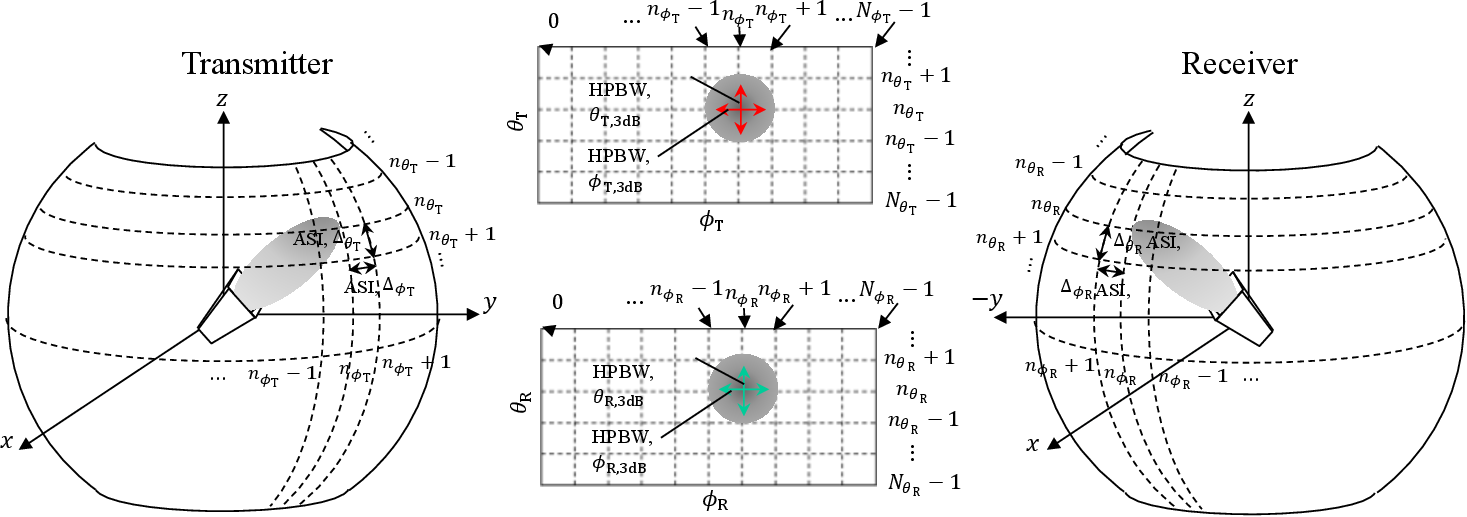}}  
\caption{Isotropic measurement vs angle-resolved measurement. \label{fig:meas_mode}}
\end{figure*}

While these methods address important aspects of directional-to-omnidirectional recovery, a practical gap remains in obtaining an accurate isotropic-equivalent received power directly from discretely sampled, non-orthogonal angular beams under realistic scan resolutions. In particular, beam overlap and sidelobe leakage bias angular power integration, and the resulting isotropic-equivalent metric further depends on the within-bin offset between the true angle and the scan grid (scalloping), which becomes increasingly pronounced for wider beams and/or coarser scan steps. Moreover, a unified framework that is straightforward to implement in various practical measurement configurations is still desirable.

This paper addresses the above gap by presenting an implementation-friendly framework for extracting
synthesized-isotropic narrowband received power, and hence path loss/path gain, from discretely sampled angle-resolved wideband measurements. The main contributions of this paper are summarized as follows:
\begin{itemize}
\item We formulate isotropic-equivalent power synthesis in a unified matrix form that explicitly accounts for the non-orthogonality and overlap of practical scan beams.
\item We introduce a beam-accumulation correction factor computed by discrete summation of the squared antenna pattern over the scan grid, together with an offset-averaged variant that mitigates scalloping without explicit off-grid angle estimation.
\item We validate the proposed framework through both channel simulations and $154$~GHz corridor measurements.
\end{itemize}

The remainder of this paper is organized as follows. Sec.~II introduces the omnidirectional narrowband and angle-resolved wideband measurement models and the discrete scan-grid representation used throughout. Sec.~III presents the synthesized-isotropic power extraction method and the proposed beam-accumulation correction, including an offset-averaged variant to mitigate scalloping. Sec.~IV describes how the accumulation factor is computed using an analytic Gaussian beam model and measured horn patterns. Sec.~V provides simulation and measurement results, and Sec.~VI concludes the paper.

\section{Measurement Methods}\label{sect:Preliminaries}

\subsection{Isotropic Narrowband Channels}
We first consider channel measurements under the ideal isotropic-antenna assumption with narrowband signaling as the traditional approach, in which neither angular nor delay resolution is available as shown in Fig.~\ref{fig:omni_meas}. As a result, the received signal represents a superposition of all multipath components (MPCs), leading to small-scale fading due to constructive and destructive interference.

Assuming isotropic antennas at both the Tx and Rx, the narrowband channel response evaluated at the carrier frequency $f_c$ is modeled as a superposition of plane-wave MPCs as
\begin{equation}
    H_0 \triangleq \sum_{l=1}^L \gamma_l e^{-\j 2\pi f_c \tau_l},
    \label{eq:H_0}
\end{equation}
where $\gamma_l$ and $\tau_l$ denote the complex amplitude and propagation delay of the $l$-th MPC, respectively, and $L$ denotes the number of MPCs.

The path gain is obtained by taking the ensemble average of the squared magnitude of the complex channel response. Expanding \eqref{eq:H_0} introduces both the power contributions of individual MPCs and cross-terms arising from interference between different paths. Under the widely accepted uncorrelated scattering (US) assumption~\cite{Bello}, the phases of different MPCs are statistically uncorrelated. Consequently, the cross-terms vanish after ensemble averaging, and the total average received power can be approximated as
\begin{equation}
    P_c = \mathbb{E}\!\left\{ H_0\cdot H_0^* \right\}
    \approx
    \sum_{l=1}^L \left| \gamma_l \right|^2,
    \label{eq:US}
\end{equation}
where $\mathbb{E}\!\left\{ \cdot \right\}$ denotes the ensemble average. This result indicates that, in a rich multipath environment with uncorrelated scatterers, the total received power is given by the incoherent sum of the powers contributed by individual MPCs.

\subsection{Angle-Resolved Wideband Channels}
We now extend the above narrowband omnidirectional model to the angle-resolved wideband channel measurement framework. Let us consider an angle-scanning measurement sampled in the angular and frequency domains with a given ASI, as illustrated in Fig.~\ref{fig:directional_scan_spherical}, and a given frequency resolution within a measurement bandwidth $W$. The frequency-domain sounding signal is assumed to be known and perfectly equalized. Hence, the influence of the transmitted waveform is removed, and the measured data represent the channel transfer function (CTF). 

The measured time-invariant angle-resolved multi-dimensional channel transfer function (MDCTF) \cite{Kim_CLEAN, Kim_Springer} is expressed as
\begin{multline}
    \tilde{H}(\check{f}, \check{\theta}_\T, \check{\phi}_\T,
    \check{\theta}_\R, \check{\phi}_\R)
    =
    \iiiint
    a_{\R}(\check{\theta}_\R-\vartheta_\R,
    \check{\phi}_\R-\varphi_\R) \\
    \cdot
    H_c(\check{f}, \vartheta_\T, \varphi_\T,
    \vartheta_\R, \varphi_\R)
    a_{\T}(\check{\theta}_\T-\vartheta_\T,
    \check{\phi}_\T-\varphi_\T) \\
    \cdot
    \sin\vartheta_\T \, d\vartheta_\T d\varphi_\T
    \sin\vartheta_\R \, d\vartheta_\R d\varphi_\R \\
    + \frac{1}{\sqrt{G_\T \, G_\R}}Z(\check{f}, \check{\theta}_\T, \check{\phi}_\T,
    \check{\theta}_\R, \check{\phi}_\R),
    \label{eq:Htilde}
\end{multline}
where $H_c(f, \vartheta_\T, \varphi_\T, \vartheta_\R, \varphi_\R)$ denotes the propagation channel transfer function in the joint frequency and angular domains. $a_\T(\cdot)$ and $a_\R(\cdot)$ denote the gain-normalized far-field radiation patterns of the Tx and Rx antennas, respectively, expressed as functions of the angular offset between the antenna pointing direction and the angle of departure or arrival. Here, $G_\T$ and $G_\R$ denote the peak (boresight) power gains of the Tx and Rx antennas, respectively, and $Z(\cdot)$ represents the additive measurement noise, modeled as zero-mean circularly symmetric complex Gaussian noise. \eqref{eq:Htilde} is written in a gain-normalized form, where the Tx/Rx antenna boresight gains are removed from the measured response and the additive noise term is scaled accordingly.

The discrete co-elevation ($\theta$), measured from the zenith direction in the adopted spherical-coordinate system, and azimuth ($\phi$) pointing angles at the Tx and Rx are defined as
\begin{align}
    \check{\theta}_\T &\in
    \left\{\theta_{\T,n_{\theta_\T}}= n_{\theta_\T}\Delta_{\theta_\T}
    \mid n_{\theta_\T}=0,\ldots,N_{\theta_\T}-1 \right\},  \\
    \check{\phi}_\T &\in
    \left\{\phi_{\T,n_{\phi_\T}}=  n_{\phi_\T}\Delta_{\phi_\T}
    \mid n_{\phi_\T}=0,\ldots,N_{\phi_\T}-1 \right\},  \\
    \check{\theta}_\R &\in
    \left\{\theta_{\R,n_{\theta_\R}}=  n_{\theta_\R}\Delta_{\theta_\R}
    \mid n_{\theta_\R}=0,\ldots,N_{\theta_\R}-1 \right\},  \\
    \check{\phi}_\R &\in
    \left\{\phi_{\R,n_{\phi_\R}}=  n_{\phi_\R}\Delta_{\phi_\R}
    \mid n_{\phi_\R}=0,\ldots,N_{\phi_\R}-1 \right\}.
\end{align}
The delay and frequency bins are given by
\begin{align}
    \check{f} &\in
    \left\{f_{n_f} =  f_c + n_f \Delta_f
    \mid n_f=-N/2,\ldots,N/2-1 \right\}, \\
    \check{\tau} &\in
    \left\{\tau_{n_\tau} = n_{\tau}\Delta_{\tau}
    \mid n_{\tau}=0,\ldots,N-1 \right\}, 
    \label{eq:freq}
\end{align}
where $f_c$ denotes the center frequency, and $n$, $N$, and $\Delta$ denote the sample index, the total number of samples, and the ASI in the corresponding domain, respectively. The delay resolution satisfies $\Delta_\tau = 1/(N\Delta_f)$. The propagation channel transfer function
\begin{align}
    H_c(f, \theta_\T,\phi_\T, & \theta_\R,\phi_\R) =
    \sum_{l=1}^{L}
    \gamma_l e^{-\j 2\pi f \tau_l} \nonumber \\
    &\quad \cdot
    \frac{1}{\sin\theta_\T} \,
    \delta(\theta_\T-\theta_{\T,l}) \,
    \delta(\phi_\T-\phi_{\T,l}) \nonumber \\
    &\quad \cdot
    \frac{1}{\sin\theta_\R} \,
    \delta(\theta_\R-\theta_{\R,l}) \,
    \delta(\phi_\R-\phi_{\R,l}),
    \label{eq:Hc}
\end{align}
which is expressed by extending \eqref{eq:H_0} to the angular domains. The factors $1/\sin\theta_\T$ and $1/\sin\theta_\R$ are included to ensure the proper normalization of the angular Dirac delta functions in spherical coordinates with surface element ($\sin\theta\,d\theta\,d\phi$). Neglecting the noise term in \eqref{eq:Htilde} (high-SNR assumption) and substituting \eqref{eq:Hc} into \eqref{eq:Htilde}, the MDCTF is obtained as
\begin{multline}
    \tilde{H}(\check{f}, \check{\theta}_\T, \check{\phi}_\T,
    \check{\theta}_\R, \check{\phi}_\R)
    \approx
    \sum_{l=1}^{L}
    \gamma_l \,
    a_\R(\check{\theta}_\R-\theta_{\R,l},
    \check{\phi}_\R-\phi_{\R,l}) \\
    \cdot
    a_\T(\check{\theta}_\T-\theta_{\T,l},
    \check{\phi}_\T-\phi_{\T,l}) \,
    e^{-\j 2\pi \check{f} \tau_l}.
    \label{eq:Htilde2}
\end{multline}
That is, the measured response can be interpreted as the underlying propagation channel weighted by the Tx/Rx directional antenna patterns at the corresponding pointing angles.

By applying the inverse Fourier transform to \eqref{eq:Htilde2}, the discrete time-invariant multi-dimensional channel impulse response (MDCIR) sample at multi-dimensional index $\Vect{n}=[n_\tau, n_{\theta_\T}, n_{\phi_\T}, n_{\theta_\R}, n_{\phi_\R}]$ is obtained as
\begin{equation}
    \tilde{h}_{\Vect{n}} = \sum_{l=1}^{L} \gamma_l A_{\Vect{n}}(\boldsymbol{\Omega}_l),
    \label{eq:hmodel_scalar}
\end{equation}
where $\Vect{\Omega}_l = [\tau_l, \theta_{\T,l}, \phi_{\T,l}, \theta_{\R,l}, \phi_{\R,l}]^T$, and the response function for the $l$-th MPC is defined as
\begin{equation}
    \begin{split}
        A_{\Vect{n}}(\boldsymbol{\Omega}_l)
        &\triangleq
        a_u(\tau_{n_\tau}-\tau_l) \\
        &\quad \cdot
        a_\R(\theta_{\R,n_{\theta_\R}}-\theta_{\R,l}, \phi_{\R,n_{\phi_\R}}-\phi_{\R,l}) \\
        &\quad \cdot
        a_\T(\theta_{\T,n_{\theta_\T}}-\theta_{\T,l}, \phi_{\T,n_{\phi_\T}}-\phi_{\T,l}).
    \end{split}
    \label{eq:Aqp}
\end{equation}
Note that due to the finite measurement bandwidth, the ideal Dirac delta function in the delay domain is replaced by the autocorrelation function of the sounding signal, denoted by $a_u(\tau)$.

By stacking all delay and angular samples into a single column vector $\tilde{\Vect{h}} \in \mathbb{C}^{N_{\tot} \times 1}$ ($N_{\tot} \triangleq N \, N_{\theta_\T} \, N_{\phi_\T}\, N_{\theta_\R}\, N_{\phi_\R}$), the discrete channel model can be equivalently expressed in vector-matrix form as
\begin{equation}
    \tilde{\Vect{h}} = \sum_{l=1}^{L} \gamma_l \Vect{A} (\Vect{\Omega}_l) = \Vect{A}\Vect{\gamma},
    \label{eq:hmodel_vec}
\end{equation}
where $\Vect{\gamma} = [\gamma_1,\dots,\gamma_L]^T$, and the matrix $\Vect{A}=[\Vect{A}(\Vect{\Omega}_1),\dots, \Vect{A}(\Vect{\Omega}_L)]\in \mathbb{C}^{N_\tot \times L}$ contains the steering vectors for all paths. The steering vector for the $l$-th path is defined using the Kronecker product as
\begin{equation}
    \Vect{A}(\boldsymbol{\Omega}_l)
    \triangleq
    \Vect{a}_u(\tau_l)
    \otimes
    \Vect{a}_\T(\theta_{\T,l},\phi_{\T,l})
    \otimes
    \Vect{a}_\R(\theta_{\R,l},\phi_{\R,l}).
    \label{eq:resfunc}
\end{equation}
The constituent vectors are defined as
\begin{align}
    \left[\Vect{a}_u(\tau_l)\right]_{n_\tau} &\triangleq  a_u(\tau_{n_\tau}-\tau_{l}) \nonumber \\
    \left[\Vect{a}_\T(\theta_{\T,l}, \phi_{\T,l})\right]_{n_{\theta_\T},n_{\phi_\T}} &\triangleq  a_\T(\theta_{\T, n_{\theta_\T}}-\theta_{\T,l}, \phi_{\T, n_{\phi_\T}} - \phi_{\T,l}),\nonumber \\
    \left[\Vect{a}_\R(\theta_{\R,l}, \phi_{\R,l})\right]_{n_{\theta_\R},n_{\phi_\R}} &\triangleq  a_\R(\theta_{\R, n_{\theta_\R}}-\theta_{\R,l}, \phi_{\R, n_{\phi_\R}} - \phi_{\R,l}). \nonumber
\end{align}

\section{Channel Parameter Calculation}
\label{sec:channel_param_calc}
\subsection{Path Gain}
The multi-dimensional channel power vector is defined as the expected element-wise squared magnitude of the MDCIR vector. Assuming the path phases are random and uncorrelated, the cross-terms in the expansion average to zero \cite{Kim_CLEAN}, namely
\begin{align}
    \Vect{P} 
    &\triangleq \mathbb{E}\!\left\{\tilde{\Vect{h}}\odot \tilde{\Vect{h}}^{*}\right\} 
    = \mathbb{E}\!\left\{(\Vect{A}\Vect{\gamma})\odot(\Vect{A}\Vect{\gamma})^{*}\right\} \nonumber\\
    &= \mathbb{E}\!\left\{\sum_{l=1}^{L} |\gamma_l|^2\,|\Vect{A}_l|^2
    + \sum_{\substack{l,l'=1\\ l'\neq l}}^{L}
    \gamma_l\gamma_{l'}^{*}
    (\Vect{A}_l\odot \Vect{A}_{l'}^{*})\right\} \nonumber \\
    &\approx \sum_{l=1}^{L} |\gamma_l|^2 \, |\Vect{A}_l|^2,
    \label{eq:power_vector_expand}
\end{align}
where $\Vect{A}_l$ denotes $\Vect{A}(\Vect{\Omega}_l)$ for simplicity, and $\odot$ denotes the Hadamard (element-wise) product. This can be rewritten in matrix notation as
\begin{align}
    \Vect{P} = \Vect{B} \, \Vect{p} \quad \in \mathbb{R}^{N_{\tot}\times 1},
    \label{eq:Bp}
\end{align}
where $\Vect{B} = |\Vect{A}|^2 = \Vect{A}\odot\Vect{A}^* \in \mathbb{R}^{N_\tot \times L}$ is the basis matrix of component-wise power responses, and $\Vect{p} = \Vect{\gamma} \odot \Vect{\gamma}^* \in \mathbb{R}^{L \times 1}$ is the vector of path powers.

For analytical and computational tractability, the continuous MPC parameter space is commonly discretized onto a fixed grid. By assuming that the channel gain within each grid cell is approximately constant and grouping the MPCs associated with each cell, \eqref{eq:Bp} is approximated as
\begin{align}
    \Vect{P} \approx \Vect{B}_g \, \Vect{p}_g \quad \in \mathbb{R}^{N_{\tot}\times 1},
    \label{eq:Bgpg}
\end{align}
where $\Vect{B}_g \in \mathbb{R}^{N_\tot \times N_\tot}$ represents the dictionary matrix of power responses defined at the grid centers, and $\Vect{p}_g \in \mathbb{R}^{N_{\tot}\times 1}$ denotes the discrete power distribution across the grid, whose elements are not necessarily obtained explicitly, unlike in MPC parameter extraction \cite{Kim_CLEAN}.

Finally, the true channel power $P_c$, defined as the sum of the powers of all MPCs, is given by
\begin{equation}
    P_c \triangleq \sum_{l=1}^{L} |\gamma_l|^2
    = \sum_{k=1}^{N_{\tot}} [\Vect{p}_g]_k
    = \Vect{1}_{N_\tot}^T \Vect{p}_g,
\end{equation}
where $\Vect{1}_{v}\in\mathbb{R}^{v\times 1}$ denotes the all-ones vector. In principle, $\Vect{p}_g$ can be obtained from \eqref{eq:Bgpg} as
\begin{equation}
    \Vect{p}_g \approx \Vect{B}_g^{-1}\Vect{P},
\end{equation}
but explicitly inverting the large matrix $\Vect{B}_g$ is computationally expensive and often unstable.

Instead, we exploit the property that the column sums of $\Vect{B}_g$ are constant. This arises from the rotational invariance of the measurement scan, where the same antenna pattern is swept across the grid. This property is expressed as
\begin{equation}
    \Vect{1}_{N_\tot}^T \Vect{B}_g = \zeta \Vect{1}_{N_\tot}^T,
    \label{eq:zeta_property}
\end{equation}
where $\zeta$ is the total beam accumulation factor. It is obtained by the sum of the elements of any column $m$ of the basis matrix as
\begin{equation}
    \zeta = \sum_{n=1}^{N_{\tot}} [\Vect{B}_g]_{n,m}.
    \label{eq:zeta}
\end{equation}
Premultiplying \eqref{eq:Bgpg} by $\Vect{1}_{N_\tot}^T$ yields
\begin{align}
    \Vect{1}_{N_\tot}^T \Vect{P}
    &= \Vect{1}_{N_\tot}^T \Vect{B}_g \Vect{p}_g \nonumber\\
    &\approx \zeta \Vect{1}_{N_\tot}^T \Vect{p}_g
    = \zeta P_c.
\end{align}
Therefore, the channel power (synthesized path gain) can be estimated directly from the measured power as
\begin{equation}
    P_c = \Vect{1}_{N_\tot}^T \Vect{p}_g
    \approx \frac{1}{\zeta}\,\Vect{1}_{N_\tot}^T \Vect{P}
    = \frac{1}{\zeta}\sum_{k=1}^{N_{\tot}} [\Vect{P}]_k,
    \label{eq:Pc_from_zeta}
\end{equation}
where $\zeta$ acts as a correction factor accounting for beam-overlap effects.

\subsection{Correction Factors}
Since the basis matrix $\Vect{B}_g$ is constructed from the Kronecker product of the delay and angular responses, the total beam accumulation factor $\zeta$ in \eqref{eq:zeta} can be factorized into the product of the contributions from each domain as
\begin{equation}
    \zeta = \zeta_\tau \cdot \zeta_\T \cdot \zeta_\R,
    \label{eq:zeta_prod}
\end{equation}
where $\zeta_\tau$, $\zeta_\T$, and $\zeta_\R$ represent the accumulation factors for the delay, Tx angle, and Rx angle domains, respectively.

The delay-domain factor is determined by the column sum of the delay basis matrix. The elements of the delay dictionary are given by
\begin{equation}
    [\Vect{B}_{g,\tau}]_{n,m} = |a_u(\tau_{n}-\tau_{m})|^2,
    \label{eq:Bg_delay_element}
\end{equation}
with $\tau_n=n\Delta_\tau$ and $\tau_m=m\Delta_\tau$ for $n,m \in \{0,\ldots,N_\tau-1\}$. For any column $m$, the accumulation factor is formally defined as
\begin{equation}
    \zeta_\tau = \sum_{n=0}^{N_\tot-1} [\Vect{B}_{g,\tau}]_{n,m},
    \label{eq:zeta_tau}
\end{equation}
where $N_\tot \triangleq N_\tau$. Since the autocorrelation function $a_u(\tau)$ depends only on the delay difference, this summation is shift-invariant (assuming a periodic grid or negligible edge effects). Therefore, \eqref{eq:zeta_tau} simplifies to the sum of the squared autocorrelation samples
\begin{equation}
    \zeta_\tau = \sum_{n=0}^{N_\tot-1} \left| a_u(n \Delta_\tau) \right|^2.    
\end{equation}

Similarly, the angular-domain factors are determined by the summation of the squared antenna pattern samples. The elements of the angular dictionary for a 2D scan are defined as
\begin{equation}
    [\Vect{B}_{g,a}]_{n,m} = \left| a(\theta_{n_\theta}-\theta_{m_\theta}, \phi_{n_\phi}-\phi_{m_\phi}) \right|^2,
    \label{eq:Bg_angle_element}
\end{equation}
where $\theta_{n_\theta} = n_\theta \, \Delta_\theta$, $\theta_{m_\theta} = m_\theta \, \Delta_\theta$, $\phi_{n_\phi} = n_\phi \, \Delta_\phi$, and $\phi_{m_\phi} = m_\phi \, \Delta_\phi$ with angular indices $n_\theta,m_\theta \in \{0,\ldots,N_\theta-1\}$ and $n_\phi,m_\phi \in \{0,\ldots,N_\phi-1\}$. The indices $n = n_\phi + n_\theta N_\phi$ and $m = m_\phi + m_\theta N_\phi$ denote the linear indices for the observation and grid points, respectively. The function $a(\cdot)$ represents the 2D antenna radiation pattern.

The angular accumulation factor is then obtained by summing the elements of any column $m$ as
\begin{equation}
    \zeta_a = \sum_{n=0}^{N_\tot - 1} [\Vect{B}_{g,a}]_{n,m},
    \label{eq:zeta_angle}
\end{equation}
where $N_\tot \triangleq N_\theta N_\phi$. The angular accumulation factor $\zeta_{a}$ in \eqref{eq:zeta_angle} is simply calculated by summing the squared magnitude of the antenna radiation pattern $a(\cdot)$ over all discrete scan angles 
\begin{equation}
    \zeta_{a}= \sum_{n_{\theta}=0}^{N_{\theta}-1} \sum_{n_{\phi}=0}^{N_{\phi}-1} \left| a(n_{\theta} \Delta_{\theta}, n_{\phi} \Delta_{\phi}) \right|^2.    
\end{equation}
Furthermore, if the 2D antenna pattern can be approximated as separable in the co-elevation (Co-El) and azimuth (Az) domains (i.e., $|a(\theta, \phi)|^2 \approx |a_{\theta}(\theta)|^2 \cdot |a_{\phi}(\phi)|^2$), the 2D correction factor can be decomposed into the product of two 1D factors as
\begin{align}
    \zeta_{a} &\approx \left( \sum_{n_{\theta}=0}^{N_{\theta}-1} |a_{\theta}(n_{\theta} \Delta_{\theta})|^2 \right) 
    \cdot 
    \left( \sum_{n_{\phi}=0}^{N_{\phi}-1} |a_{\phi}(n_{\phi} \Delta_{\phi})|^2 \right)\\
    &= \zeta_{\theta} \cdot \zeta_{\phi},
    \label{eq:zeta_ang_sep}
\end{align}
where $\zeta_{\theta}$ and $\zeta_{\phi}$ represent the beam accumulation factors for the Co-El and Az scans, respectively.

Based on this formulation, the correction factors for various measurement configurations are derived as follows.

\subsubsection{Omnidirectional Wideband Case}
In this case, isotropic antennas are assumed ($\zeta_\T = \zeta_\R = 1$), and the system provides only delay resolution. The autocorrelation function of the sounding signal is defined as the Dirichlet kernel \cite{Kim_Springer} as
\begin{equation}
    a_u(\tau) = \frac{1}{N} e^{-\mathrm{j}\pi \Delta_f \tau} \frac{\sin(\pi N \Delta_f \tau)}{\sin(\pi \Delta_f \tau)}.
    \label{eq:autau}
\end{equation}
For a standard wideband measurement where the delay grid matches the sampling interval ($\Delta_\tau = 1/W$), the orthogonality of the Dirichlet kernel implies that the delay accumulation factor is unity ($\zeta_\tau = 1$). Consequently, the total correction factor simplifies to
\begin{equation}
    \zeta = \zeta_\tau \cdot \zeta_\T \cdot \zeta_\R = 1 \cdot 1 \cdot 1 = 1.
\end{equation}
Thus, no correction is required, and the total channel power is obtained simply by summing the samples of the power delay profile.

\subsubsection{AoA-Resolved Narrowband Case}
In this case, the system employs angular scanning at the receiver but lacks delay resolution ($\zeta_\tau=1$). Assuming an isotropic Tx ($\zeta_\T=1$), the total factor is determined solely by the Rx's beam overlap as
\begin{equation}
    \zeta = \zeta_\R \approx \zeta_{\theta_\R} \, \zeta_{\phi_\R}.
\end{equation}

\subsubsection{AoD-Resolved Narrowband Case}
Similarly, if angular scanning is performed only at the Tx with an isotropic Rx ($\zeta_\R=1$), the correction factor becomes
\begin{equation}
    \zeta = \zeta_\T \approx \zeta_{\theta_\T} \, \zeta_{\phi_\T}.
\end{equation}

\subsubsection{Double-Directional Angle-Resolved Case}
If angular scanning is performed both at the Tx and Rx, the correction factor becomes
\begin{equation}
    \zeta = \zeta_\T \, \zeta_\R \approx \zeta_{\theta_\T} \, \zeta_{\phi_\T} \, \zeta_{\theta_\R} \, \zeta_{\phi_\R}.
\end{equation}

\subsection{Delay Spread}
For PL evaluation and comparison with conventional omnidirectional wideband measurements, it is often useful to collapse the angle-resolved multidimensional power vector $\Vect{P}\in\mathbb{R}^{N_{\tot}\times 1}$ into a delay-only power delay profile (PDP) vector.
Let $N_\tau$ denote the number of delay bins, and define the number of angular scan points at the Tx and Rx as
\begin{equation}
    N_{\T} \triangleq N_{\theta_\T}N_{\phi_\T}, \qquad
    N_{\R} \triangleq N_{\theta_\R}N_{\phi_\R},
\end{equation}
such that $N_{\tot} = N_\tau N_{\R} N_{\T}$.
Using the same stacking order implied by the Kronecker-structured response model in \eqref{eq:resfunc}, the angle-collapsed (synthesized) PDP vector $\Vect{P}_\tau\in\mathbb{R}^{N_\tau\times 1}$ is obtained by summing over all Tx/Rx scan angles and compensating for the composite beam-overlap gain in the angular domains:
\begin{equation}
    \Vect{P}_{\tau}
    \triangleq
    \frac{1}{\zeta_{\T}\zeta_{\R}}
    \left(
    \Vect{I}_{N_{\tau}}
    \otimes
    \Vect{1}_{N_{\R}}^{T}
    \otimes
    \Vect{1}_{N_{\T}}^{T}
    \right)\Vect{P},
    \label{eq:Ptau_from_P}
\end{equation}
where $\Vect{I}_{N_\tau}$ is the $N_\tau\times N_\tau$ identity matrix.

If the antenna is mounted such that its center of rotation coincides with its phase center, the angular scan does not introduce any geometry-induced path-length variation (i.e., phase center eccentricity). In this ideal configuration, steering the antenna only changes the angular weighting of the incoming MPCs based on the antenna radiation pattern. Consequently, the beam-overlap correction factor (e.g., $1/(\zeta_{\T}\zeta_{\R})$ for the D-D case) acts as a global scalar multiplier on the synthesized omnidirectional PDP and does not modify its delay-domain shape.

Since the RMS delay spread (DS) is computed from the normalized PDP as
\begin{equation}
    \sigma_\tau = \sqrt{ \frac{\sum_{n} \tau_n^{2} [\Vect{P}_\tau]_n}{\sum_{n} [\Vect{P}_\tau]_n} - \left( \frac{\sum_{n} \tau_n [\Vect{P}_\tau]_n}{\sum_{n} [\Vect{P}_\tau]_n} \right)^2 },
    \label{eqn:ds}
\end{equation}
any constant scaling factor applied to $\Vect{P}_\tau$ cancels out in the numerator and denominator. Therefore, the DS estimate is invariant to the beam-overlap correction. It follows that the correction factor is critical for the accuracy of the absolute path gain but does not impact the delay dispersion characteristics.

\section{Beam Accumulation Factor for Antenna Patterns}
\subsection{Analytic Model}
In this study, a simplified 2D Gaussian beam pattern is constructed utilizing separable Von Mises functions \cite{Kim_AWPL,Corridor300GHz}\footnote{A slightly modified von-Mises-based formulation is used to facilitate HPBW tuning (i.e., to match a prescribed HPBW more directly).} for the Co-El ($\theta$) and Az ($\phi$) components. The total radiation pattern is expressed as:
\begin{equation}
    a(\theta, \phi) = a_{\theta}(\theta) \, a_{\phi}(\phi),
\label{eq:vmbeam}
\end{equation}
where the directional components for a given angular dimension $\{\theta, \phi\}$ are modeled respectively as
\begin{equation}
  a_{\theta}(\theta) = e^{\kappa_\theta(\cos\theta-1)}, \quad a_{\phi}(\phi) = e^{\kappa_\phi(\cos\phi-1)},
\label{eq:vmbeam_1D}
\end{equation}
which denote the normalized Von Mises functions (peak unity), and we use the power gain $G(\kappa_\theta,\kappa_\phi)$ calculated as
\begin{equation}
    G(\kappa_\theta,\kappa_\phi) = G_\theta(\kappa_\theta)\,G_\phi(\kappa_\phi),
\end{equation}
where
\begin{equation}
    G_\theta(\kappa_\theta) = \frac{e^{\kappa_\theta}}{I_0(\kappa_\theta)}, \quad G_\phi(\kappa_\phi) = \frac{e^{\kappa_\phi}}{I_0(\kappa_\phi)}.
\end{equation}
The concentration parameters $\kappa_\theta$ and $\kappa_\phi$ are determined by the half-power beamwidths (HPBWs), $\theta_\threedB$ and $\phi_\threedB$, as 
\begin{equation}
    \kappa_\theta = \frac{\ln \sqrt{2}}{1-\cos(0.5 \theta_\threedB)}, \quad
    \kappa_\phi = \frac{\ln \sqrt{2}}{1-\cos(0.5 \phi_\threedB)},
\end{equation}
where $I_m(\cdot)$ denotes the modified Bessel function of the first kind of order $m$. This formulation ensures that the generated pattern satisfies the specified HPBW requirements in the principal planes. 

\begin{figure}[t]
\centering
\includegraphics[width=.9\linewidth]{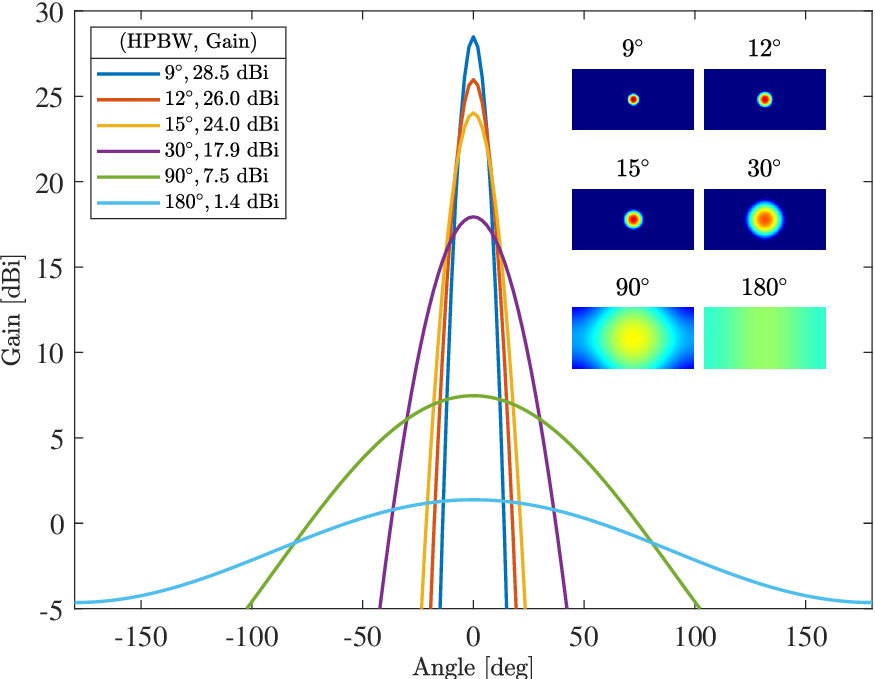}
\caption{Radiation patterns of Gaussian beams given by $G \cdot |a(\theta, \phi)|^2$ ($\phi_\threedB=\theta_\threedB$).}
\label{fig:Gaussian_pattern}
\end{figure}

The von-Mises-based Gaussian beam model is adopted not only as a simple main-lobe approximation but also for its analytical tractability \cite{Kim_AWPL}. Because the von Mises kernel admits a Fourier--Bessel expansion, key quantities required for power synthesis can be derived in closed or semi-closed forms. This enables fast evaluation and parameter sweeps over HPBW and ASI without resorting to heavy numerical integration. Fig.~\ref{fig:Gaussian_pattern} depicts the Az radiation patterns of the Gaussian beam models $G\cdot|a(\theta,\phi)|^2$ at $\theta=$\Ang{90}, configured with various HPBWs. In each angular dimension, an omnidirectional (flat) response corresponds to a beamwidth that spans the entire scan domain. Thus, the omni condition is $\phi_{\threedB}=360^\circ$ for Az ($\phi\in[0,360^\circ)$) and $\theta_{\threedB}=180^\circ$ for Co-El ($\theta\in[0,180^\circ]$). In implementation, we set $\kappa_\theta=0$ and $\kappa_\phi=0$ when $\theta_{\threedB}$ and $\phi_{\threedB}$ are equal to or larger than the corresponding domain, yielding $a_\theta(\theta)=1$ and $a_\phi(\phi)=1$, respectively.

As established in Sec.~\ref{sec:channel_param_calc}, the synthesized normalization relies on the angular accumulation factor $\zeta$, defined as the sum of the squared antenna pattern samples over the discrete scan grid. By leveraging the separability of the directional antenna model, the 2D angular accumulation factor factorizes into 1D components. In this study, we avoid the continuous-domain approximation and evaluate the accumulation factor directly on the discrete scan grid, since the scan interval $\Delta_\psi$ is not necessarily fine compared to the antenna HPBW.

Substituting the antenna model into the definition of the 1D accumulation factor yields the discrete form
\begin{equation}
\zeta_{\psi}
= \sum_{n=0}^{N_\psi-1} e^{2 \kappa_\psi \left(\cos(n\Delta_\psi)-1 \right)},
\label{eq:zeta_1d_discrete}
\end{equation}
which is computed numerically for the given scan grid $\{n\Delta_\psi\}_{n=0}^{N_\psi-1}$.

For uniform full-Az scanning, i.e., $\Delta_\psi=360^\circ/N_\psi$, \eqref{eq:zeta_1d_discrete} can be rewritten exactly using the Fourier--Bessel expansion of the von Mises kernel as
\begin{equation}
\zeta_{\psi}
= \frac{N_\psi}{e^{2\kappa_\psi}}
\left[
I_0(2\kappa_\psi)
+ 2\sum_{k=1}^{\infty} I_{kN_\psi}(2\kappa_\psi)
\right].
\label{eq:zeta_1d_discrete_bessel}
\end{equation}
In practice, \eqref{eq:zeta_1d_discrete} is evaluated directly; alternatively, \eqref{eq:zeta_1d_discrete_bessel} provides an exact reference expression whose series can be truncated when higher-order terms become negligible.

\subsection{Scalloping Loss Mitigation}
The accumulation factor in \eqref{eq:zeta_angle} is constant across grid columns when the MPC angles coincide with the angular grid centers. In practice, the true angles of MPCs are continuous and generally off-grid, which leads to a grid-dependent gain variation (scalloping loss) with respect to the within-bin angular offset.
To incorporate this effect without explicitly estimating the offset, we use the within-bin average accumulation factor.

For a 1D scan in an angular dimension $\psi\in\{\theta,\phi\}$ with scan interval $\Delta_\psi$ and $N_\psi$ scan points, let $\delta_\psi\in[0,\Delta_\psi)$ denote the angular offset from the nearest grid point. The scalloping-loss-aware 1D accumulation factor is defined as
\begin{equation}
\zeta_{\psi}(\delta_\psi)
\triangleq
\sum_{n=0}^{N_\psi-1}
\left|a_\psi(n\Delta_\psi-\delta_\psi)\right|^2.
\label{eq:zeta_1d_delta_def}
\end{equation}
Specifically, for a scan interval $\Delta_\psi$, we average the discrete-grid accumulation over a uniformly
distributed offset $\delta\sim\mathcal{U}[0,\Delta_\psi)$ as
\begin{equation}
\bar{\zeta}_\psi \triangleq \frac{1}{\Delta_\psi}\int_{0}^{\Delta_\psi} \zeta_\psi(\delta)\,d\delta.
\label{eq:zeta_delta_avg}
\end{equation}
In numerical implementation, \eqref{eq:zeta_delta_avg} is obtained by sampling $\delta$ densely over $[0,\Delta_\psi)$ and averaging the corresponding $\zeta_\psi(\delta)$ values.

Using the von-Mises Gaussian beam in \eqref{eq:vmbeam_1D}, the offset-dependent accumulation factor defined in
\eqref{eq:zeta_1d_delta_def} can be rewritten by applying the
Fourier--Bessel expansion of the von Mises kernel as
\begin{equation}
\zeta_{\psi}(\delta)=\frac{N_{\psi}}{e^{2\kappa_\psi}}
\Big[I_{0}(2\kappa_\psi)+2\sum_{m=1}^{\infty}I_{mN_{\psi}}(2\kappa_\psi)\cos(mN_{\psi}\delta)\Big].
\label{eq:zeta_psi_delta_series}
\end{equation}
Taking the within-bin average over $\delta\in[0,\Delta_{\psi})$ removes all cosine terms, yielding the closed-form expression as
\begin{equation}
\bar{\zeta}_{\psi}=\frac{N_{\psi}}{e^{2\kappa_\psi}}\,I_{0}(2\kappa_\psi).
\label{eq:zeta_psi_bar_closed}
\end{equation}
\begin{figure}[t]
\centering
\includegraphics[width=0.9\linewidth]{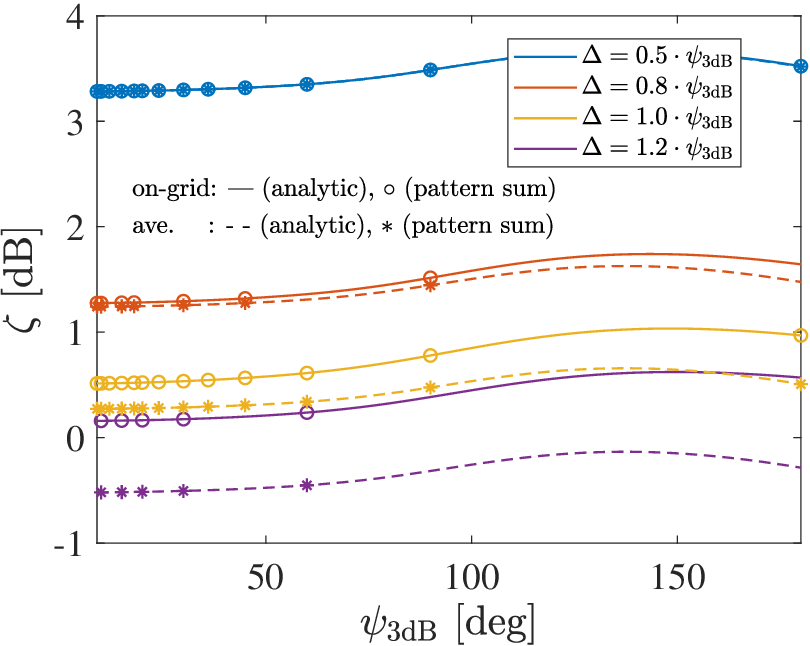}
\caption{Variation of the 1D beam-accumulation factor versus HPBW and ASI for Gaussian beam models. The on-grid accumulation factor is computed by the discrete summation in \eqref{eq:zeta_1d_discrete} (markers) and by the exact series expression in \eqref{eq:zeta_1d_discrete_bessel} for uniform full-azimuth scans (solid lines). Furthermore, the within-bin averaged factor is given by \eqref{eq:zeta_delta_avg} (markers) and by the exact series expression in \eqref{eq:zeta_psi_delta_series} (dashed lines). }
\label{fig:zeta_vmGaussian}
\end{figure}

Fig.~\ref{fig:zeta_vmGaussian} shows the 1D beam-accumulation factor as a function of HPBW and ASI. The solid curves are computed from the exact series in \eqref{eq:zeta_1d_discrete_bessel} (uniform full-azimuth scan), while the markers are obtained by the direct discrete summation in \eqref{eq:zeta_1d_discrete}. The markers are plotted only when the number of scan samples is an integer, i.e., $\Delta_\psi=360^\circ/N_\psi$ with integer $N_\psi$, to ensure the uniform periodic grid assumed in \eqref{eq:zeta_1d_discrete_bessel}. The results confirm that $\zeta_\psi$ is mainly governed by the sampling density relative to the beamwidth: smaller $\Delta_\psi/\psi_\threedB$ increases $\zeta_\psi$ due to stronger beam overlap. This trend highlights that $\zeta_\psi$ is primarily governed by the relative sampling density with respect to the beamwidth, rather than the beamwidth alone. Fig.~\ref{fig:zeta_vmGaussian} also presents the within-bin averaged factor $\bar{\zeta}_\psi$ in \eqref{eq:zeta_delta_avg} as a scalloping-robust correction.

\begin{figure}[t]
\centering
\includegraphics[width=.9\linewidth]{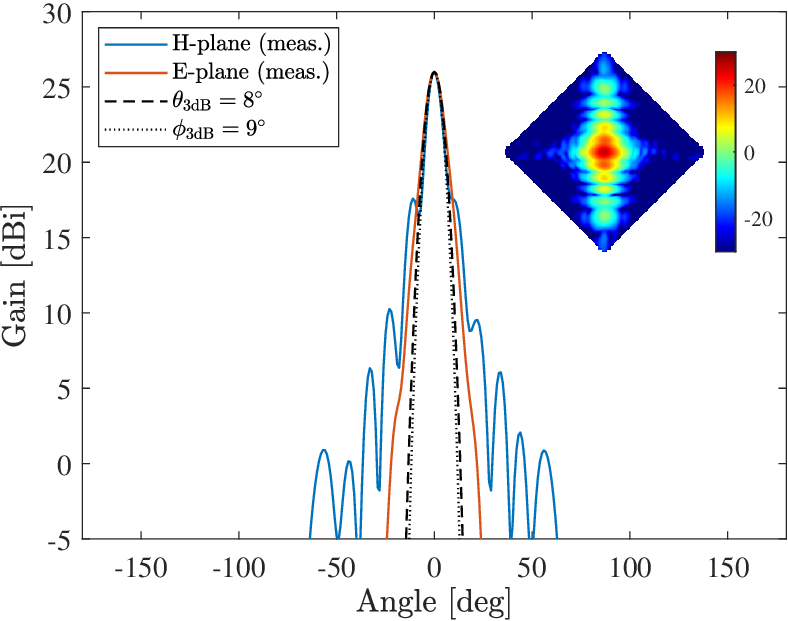}
\caption{Radiation patterns of pyramidal horn antenna ($G=26$ dBi, $\theta_\mathrm{3dB}\approx 8^\circ, \phi_\mathrm{3dB}\approx 9^\circ$) \cite{Corridor300GHz}.}
\label{fig:Horn_patterns}
\end{figure}
\begin{figure}[t]
\centering
\includegraphics[width=\linewidth]{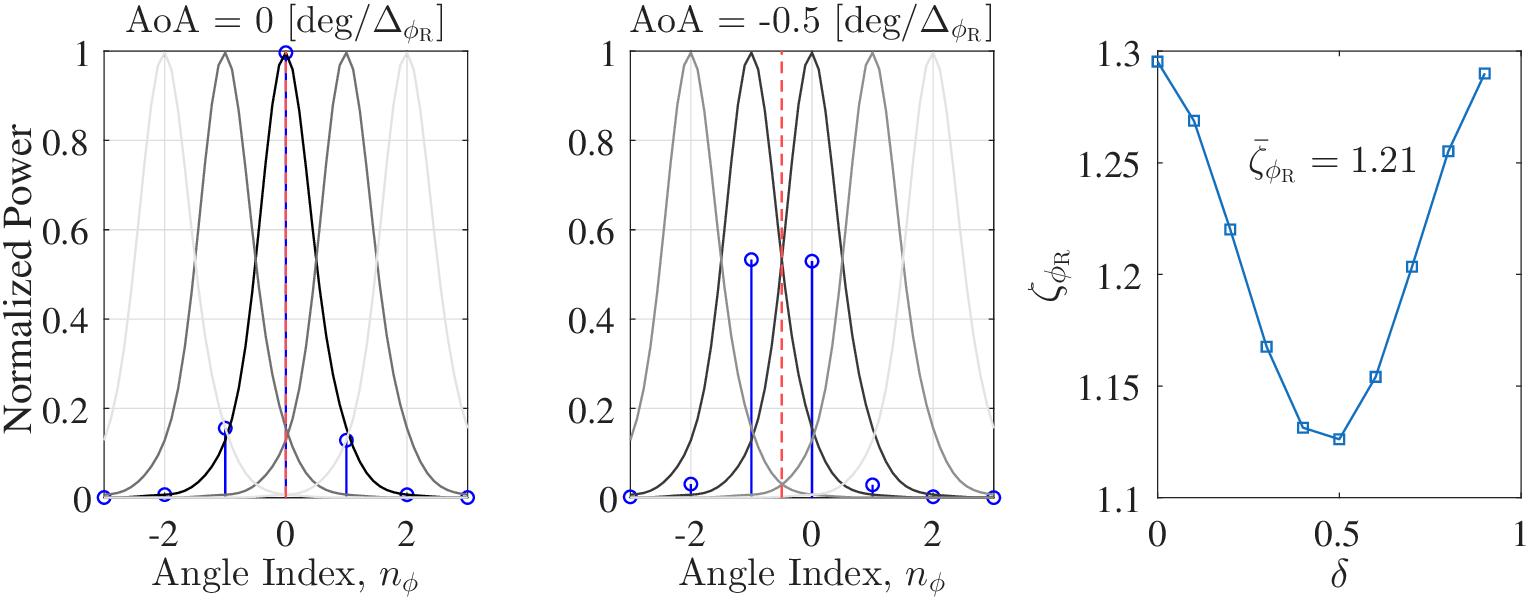}
\caption{Illustration of scalloping loss and offset-dependent beam accumulation: normalized received power versus angular index for the on-grid and half-bin-offset AoA cases, and the resulting accumulation factor $\zeta_{\phi_\R}(\delta)$ as a function of the within-bin angular offset $\delta$; the averaged value is $\bar{\zeta}_{\phi_\R}=1.21$ ($0.83$ [dB]).}
\label{fig:zeta_horn_ave}
\end{figure}

\subsection{Measured Antenna Patterns}
To evaluate the beam accumulation factor, we employ measured radiation patterns of a pyramidal horn antenna (WR6.5, $154$~GHz) \cite{Corridor300GHz,NU_SC} obtained from a 2D angular scan. Fig.~\ref{fig:Horn_patterns} compares the measured H-plane and E-plane cuts with the Gaussian beam profiles using the same HPBWs. The measured and modeled patterns agree well in the main-lobe region, while the measured E-plane exhibits noticeable sidelobes that are not captured by the Gaussian model.

Fig.~\ref{fig:zeta_horn_ave} provides an intuitive Az scanning example of scalloping in practical directional scans using the measured horn pattern in linear scale. The normalized received power (left) is shown for two AoA cases: an on-grid arrival and a half-bin offset arrival, which yields a lower peak response due to the within-bin angular mismatch. The corresponding within-bin accumulation factor $\zeta_{\phi_\R}(\delta)$ (right) varies with the offset $\delta\in[0,\Delta_\phi)$, and its average over $\delta$ gives $\bar{\zeta}_{\phi_\R}=1.21$, which is used as the offset-averaged correction in the subsequent PL synthesis.
\begin{figure}[t]
\centering
\includegraphics[width=.9\linewidth]{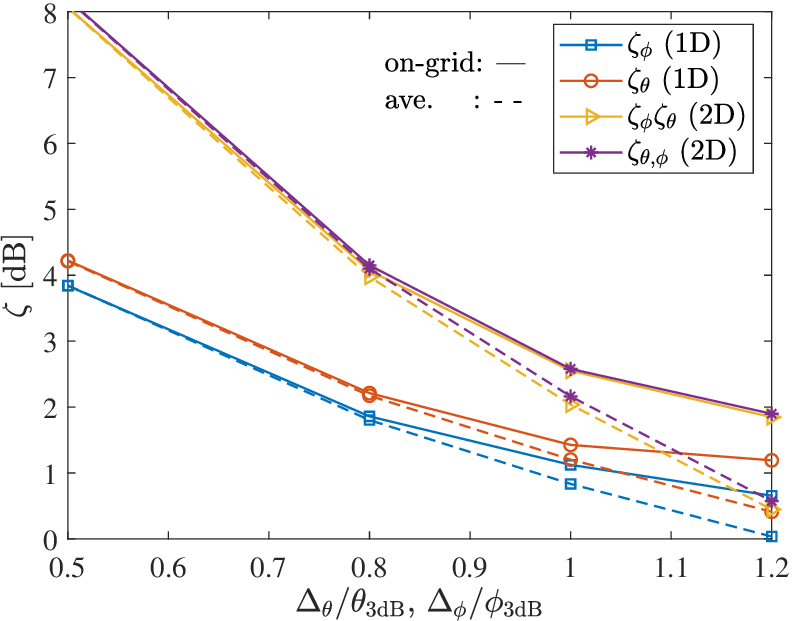}
\caption{Beam-accumulation factor of the horn antenna as a function of the ASI.}
\label{fig:zeta_Horn}
\end{figure}
\begin{table}[t]
    \centering
    \caption{Correction factors ($\zeta_{\mathrm{tot}}$) for various configurations using the horn antennas shown in Fig.~\ref{fig:Horn_patterns}.}
    \label{tab:correction_factors}
    \begin{tabular}{ccccc}
        \toprule
        \multicolumn{2}{c}{Scan domain} & \multicolumn{3}{c}{Correction factor} \\
        \cmidrule(lr){1-2} \cmidrule(lr){3-5}
        Tx & Rx & $\zeta_\tot$ & On-grid [dB] & Ave. [dB] \\
        \midrule
        Az-omni   & Az  & $\zeta_{\phi_\R}$  & $1.12$ & $0.83$ \\
        El-omni   & Co-El  & $\zeta_{\theta_\R}$   & $1.43$ & $1.21$ \\
        Isotropic & Az, Co-El & $\zeta_{\phi_\R}\zeta_{\theta_\R}$ & $2.55$ & $2.04$ \\
        Az        & Az     & $\zeta_{\phi_\R}\zeta_{\phi_\R}$ & $2.25$ & $1.67$ \\
        Az, El    & Az, Co-El & $\zeta_{\phi_\T}\zeta_{\theta_\T}\zeta_{\phi_\R}\zeta_{\theta_\R}$ & $5.10$ & $4.08$ \\
        \bottomrule
    \end{tabular}
\end{table}

Fig.~\ref{fig:zeta_Horn} further shows the beam accumulation factor computed directly from the measured horn patterns as a function of the ASI, expressed in a normalized form relative to the measured HPBW. Consistent with the analytic observation in Fig.~\ref{fig:zeta_vmGaussian}, the accumulation factor decreases as the scan becomes coarser, because the beam overlap between adjacent pointing directions diminishes. This result provides a practical lookup for selecting the appropriate correction factor under a given scan grid when measured antenna patterns are available.

Table~\ref{tab:correction_factors} lists the total beam-accumulation correction factors $\zeta_{\mathrm{tot}}$ for representative scan configurations using the measured horn patterns in Fig.~\ref{fig:Horn_patterns}. 
Here, $\zeta_{\mathrm{tot}}$ is given by the product of the relevant 1D/2D accumulation factors in the scanned domains (e.g., Az/Co-El at Tx/Rx), while the last two columns report $\zeta_{\mathrm{tot}}$ in dB for the on-grid and within-bin averaged corrections, respectively.

\section{Evaluations}
Although the correction factor has been formulated, its practical value must be verified by showing how accurately the synthesized power can be recovered under realistic multipath conditions. \revision{yellow}{(R1-1)}{}{To this end, we validate the proposed method in two complementary ways: (i) by simulation, where the true channel power is known and the estimation error can be directly quantified, and (ii) by measurement, where three independently obtained PL estimates are compared to assess their mutual consistency.}

\begin{figure*}[t]
\begin{center}
\subfigure[HPBW: \Ang{15}.\label{fig:Spectra_15deg}]{ \includegraphics[width=0.45\linewidth]{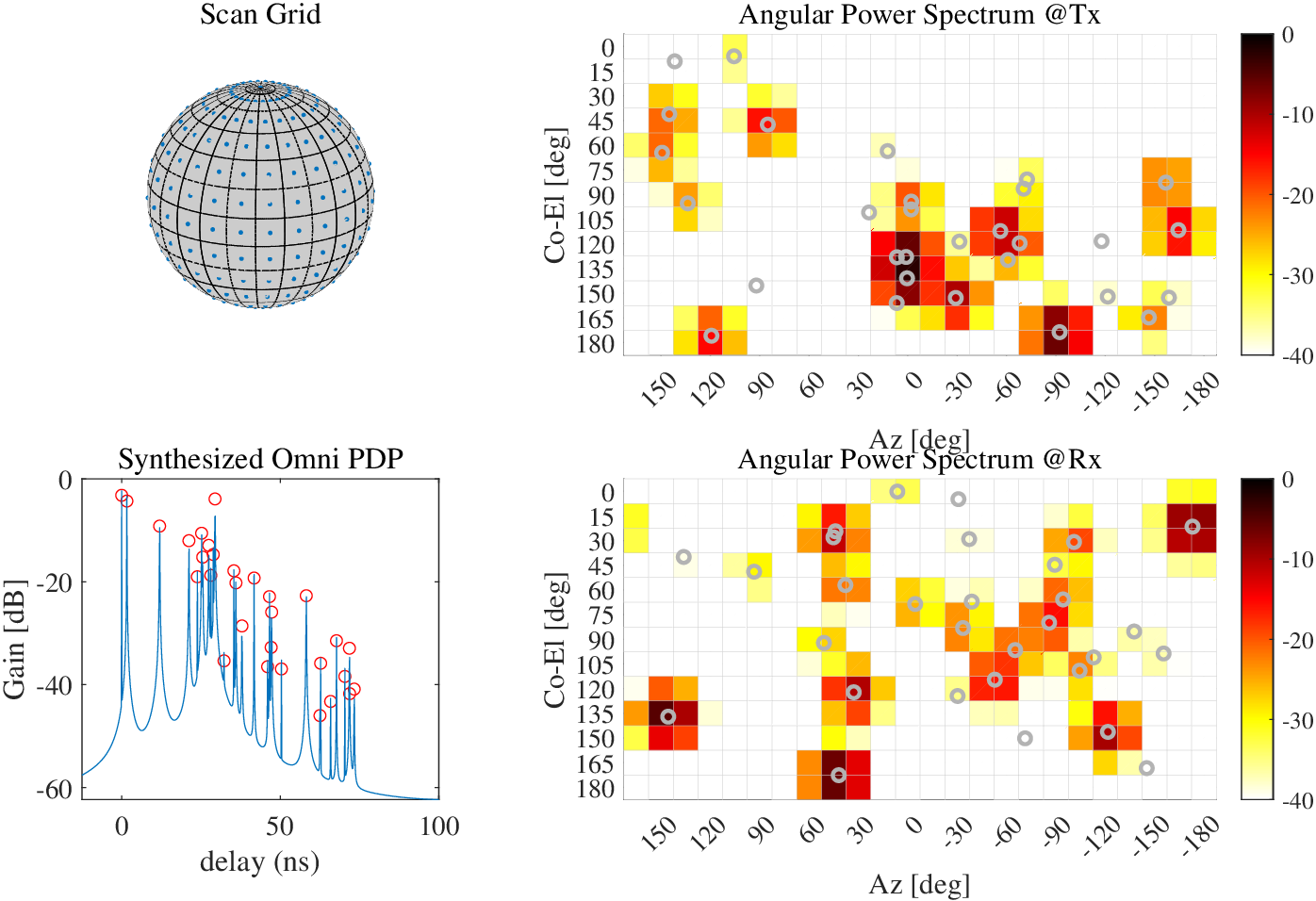}}
\qquad 
\subfigure[HPBW: \Ang{30}.\label{fig:Spectra_30deg}]{ \includegraphics[width=0.45\linewidth]{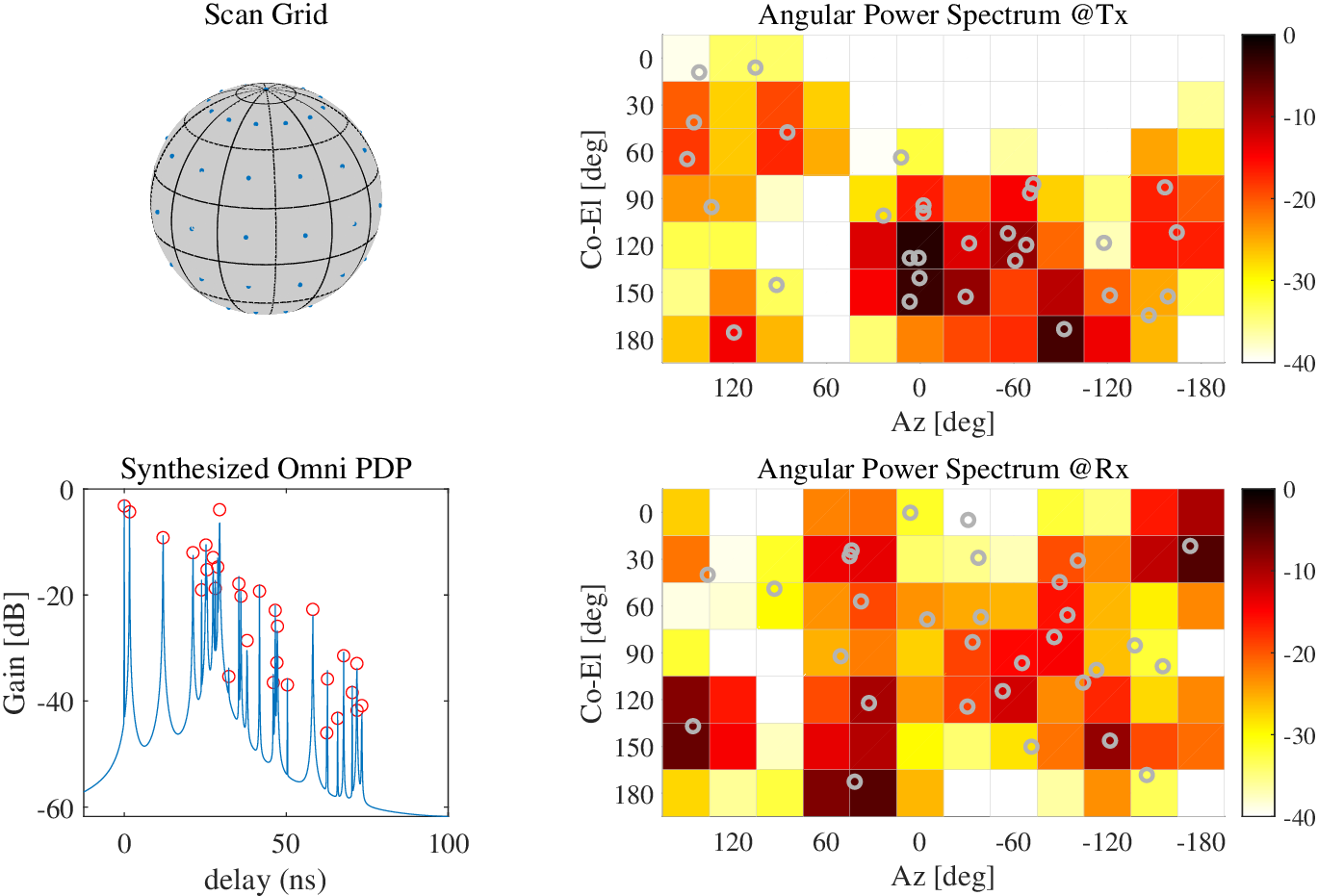}}\\
\subfigure[HPBW: \Ang{60}.\label{fig:Spectra_60deg}]{ \includegraphics[width=0.45\linewidth]{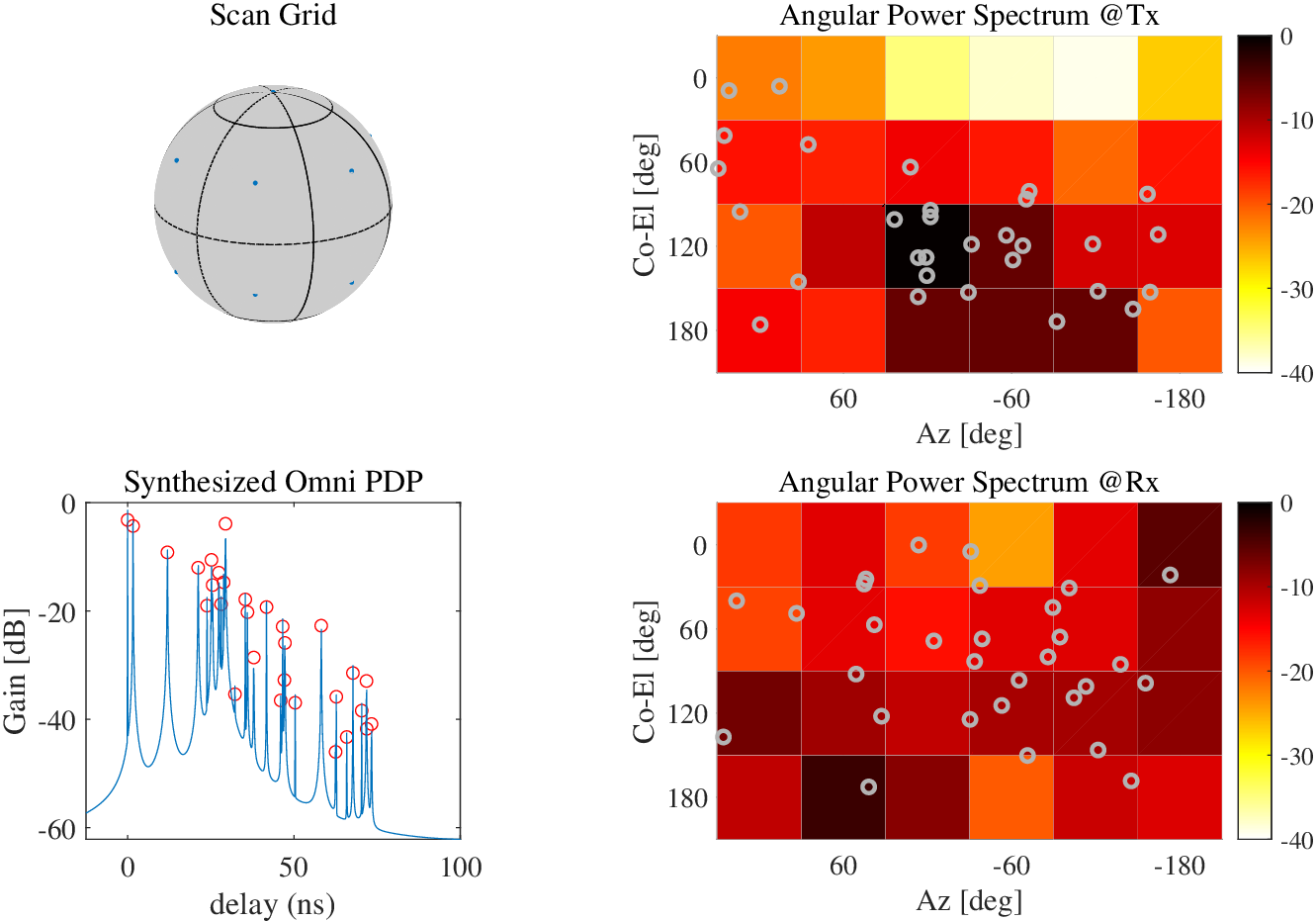}}
\qquad 
\subfigure[Isotropic.\label{fig:Spectra_360deg}]{ \includegraphics[width=0.45\linewidth]{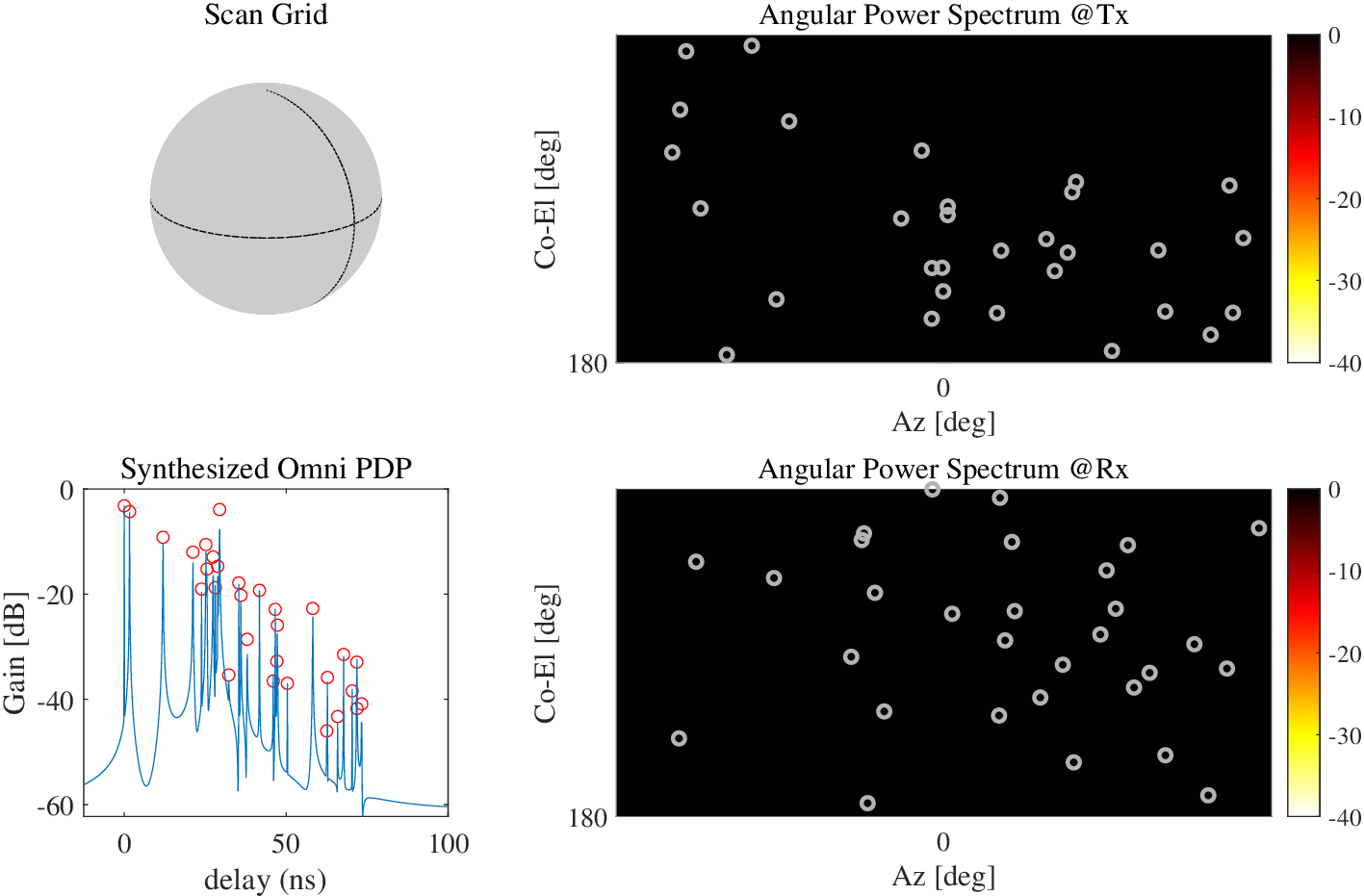}}
\caption{Synthesized power spectra as a function of the antenna HPBW, assuming identical HPBWs in Co-El and Az at both Tx and Rx
($\theta_{\T,\threedB}=\phi_{\T,\threedB}=\theta_{\R,\threedB}=\phi_{\R,\threedB}$), where the ASI is equal to the HPBW. The markers indicate the MPCs in delay and AoD/AoA angle domains. \label{fig:Spectra}}
\end{center}
\end{figure*}

\subsection{Simulation Using a Channel Model}
\label{sec:simbasedeval}
\subsubsection{Channel Model}
The underlying propagation channel is generated using a classical Saleh--Valenzuela (SV)-type clustered multipath model \cite{SV}. Cluster arrivals follow a Poisson process with rate $\Lambda$, and intra-cluster ray arrivals follow a Poisson process with rate $\lambda$. Let $T_k$ denote the arrival time of the $k$-th cluster, and $\tau_{k,l}$ denote the absolute delay of the $l$-th ray within the $k$-th cluster. The average power of the $l$-th ray in the $k$-th cluster decays exponentially in both the cluster and ray levels as
\begin{equation}
\mathbb{E}\{P_{k,l}\}
= \Xi_k \exp\!\left(-\frac{T_k}{\varrho}\right)\exp\!\left(-\frac{\tau_{k,l}-T_k}{\rho}\right),
\label{eq:sv_power_decay}
\end{equation}
where $\varrho$ and $\rho$ are the cluster- and ray-level decay constants, respectively, and $\Xi_k$ models cluster-to-cluster shadowing by a lognormal factor. Each ray is assigned a complex gain
\begin{equation}
\gamma_{k,l}=\sqrt{P_{k,l}}\,\nu_{k,l},\qquad \nu_{k,l}\sim \mathcal{CN}(0,1),
\label{eq:sv_complex_gain}
\end{equation}
and an additional independent random phase can be applied in each Monte-Carlo trial to enforce uncorrelated scattering across MPCs.

For the angular domain, each cluster is assigned a mean departure/arrival direction. The cluster mean Az angles are drawn uniformly over $[0,2\pi)$, while the cluster mean zenith (co-elevation) angles are concentrated around broadside (near $\theta\approx 90^\circ$). Individual ray angles are generated by adding independent Laplacian-distributed offsets to the corresponding cluster mean angles as
\begin{equation}
\phi = \bar{\phi}_k + \Delta\phi, \quad \mathrm{and} \quad
\theta = \bar{\theta}_k + \Delta\theta,
\label{eq:sv_angle_offsets}
\end{equation}
where $\Delta\phi$ and $\Delta\theta$ follow Laplacian distributions whose spreads are controlled by the prescribed Az/Co-El angular-spread (AS) parameters.

\begin{figure*}[t]
\begin{center}
\subfigure[Az Scan@Rx.\label{fig:AzRx}]{ \includegraphics[width=0.45\linewidth]{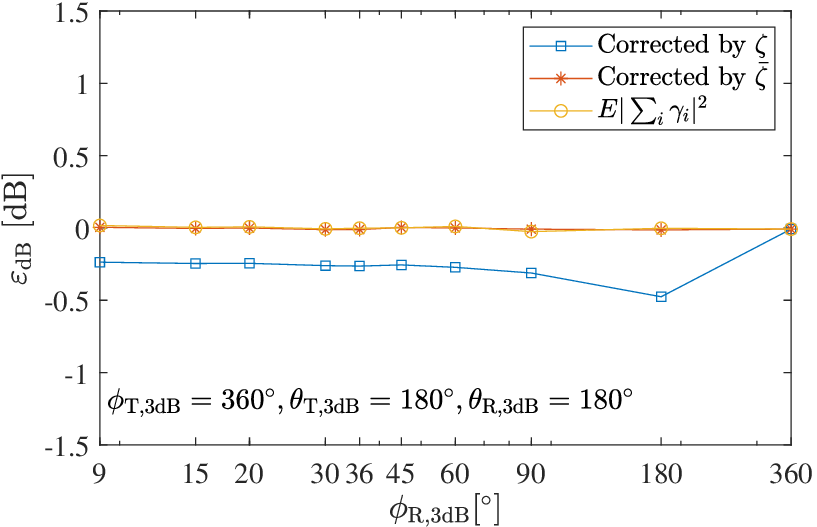}}
\qquad 
\subfigure[Co-El Scan@Rx.\label{fig:ElRx}]{ \includegraphics[width=0.45\linewidth]{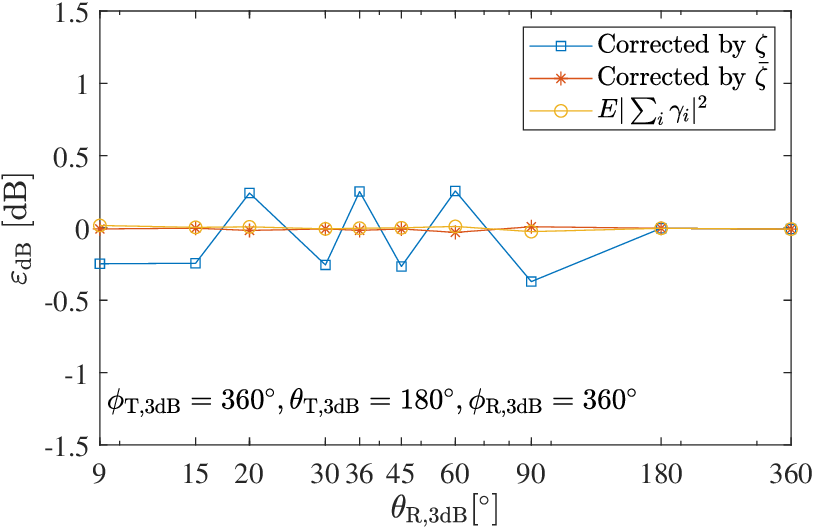}}\\
\subfigure[Az Scan@Tx,Rx.\label{fig:Az2TxRx}]{ \includegraphics[width=0.45\linewidth]{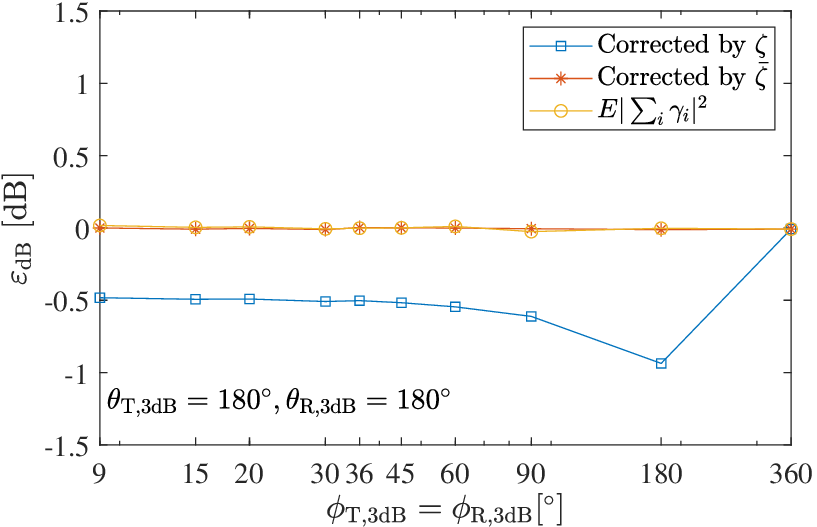}}
\qquad 
\subfigure[Co-El,Az Scan@Tx,Rx.\label{fig:AzElTxRx}]{ \includegraphics[width=0.45\linewidth]{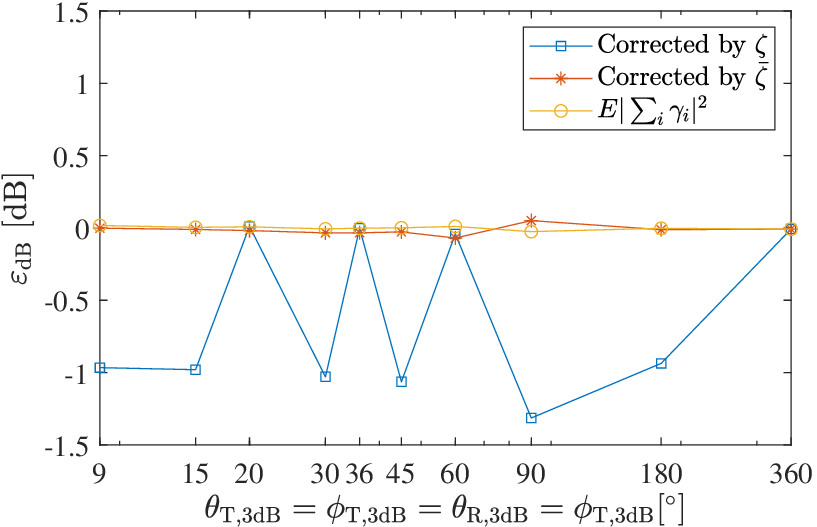}}
\caption{Path gain estimation error versus HPBW for various directional scanning configurations, evaluated relative to the incoherent power sum $P_c=\sum |\gamma_l|^2$. \label{fig:estimation_error}}
\end{center}
\end{figure*}

\subsubsection{Simulation Parameters}
We set the channel model parameters as $\varrho = 10~\mathrm{ns}$ and $\rho = 5~\mathrm{ns}$, which implies $\Lambda = 1/\varrho$ and $\lambda = 1/\rho$. Hence, the mean inter-arrival times are $1/\Lambda=\varrho$ for clusters and $1/\lambda=\rho$ for rays, controlling the delay-domain sparsity and the number of effective MPCs. To stress-test the proposed power synthesis and the beam-accumulation correction under a highly diffuse condition, we also consider a wide-spread case with $(\sigma_{\phi}, \sigma_{\theta}) = (100^\circ, 100^\circ)$. This wide-spread configuration increases the probability that MPCs populate the full angular domain and is effective for evaluating residual grid-dependence (e.g., HPBW-dependent bias), particularly in Co-El scanning. The cluster-to-cluster large-scale variability is modeled by a lognormal shadowing term with standard deviation $\sigma_{\mathrm{cl}}=3~\mathrm{dB}$ applied to the cluster power level. This term emulates random fluctuations of dominant clusters across realizations and avoids unrealistically deterministic cluster amplitudes. Assuming sub-terahertz communications, we set the bandwidth to $4$~GHz (delay bin resolution is $250$~ps).

\subsubsection{Evaluation Results}
\label{subsec:eval_proc}
For a given antenna HPBW setting, we first construct the discrete scan grids in delay and angle, and generate a clustered multipath channel realization according to \eqref{eq:sv_power_decay}--\eqref{eq:sv_angle_offsets}. Using the resulting MPC parameters $\{ \tau_l, \theta_{\T,l}, \phi_{\T,l}, \theta_{\R, l}, \phi_{\R, l}, \gamma_l\}_{l=1}^{L}$, we synthesize the angle-resolved wideband measurement by sampling the multi-dimensional channel response on the prescribed scan grid, consistent with the response model in \eqref{eq:hmodel_scalar}--\eqref{eq:resfunc}. 

Fig.~\ref{fig:Spectra} shows representative synthesized PDP and angular power spectra (APS) obtained from the same underlying multipath channel realization under different antenna HPBWs (with identical HPBWs in Co-El and Az at both Tx and Rx). The purpose of this figure is to illustrate that the observed synthesized responses vary with antenna beamwidth even when the underlying channel is identical. For narrow beams (e.g., ${\rm HPBW}=15^\circ$), the spectrum exhibits sharp angular selectivity and limited leakage to adjacent bins. As the HPBW increases (e.g., ${\rm HPBW}=30^\circ$ and $60^\circ$), the beam overlap and sidelobe leakage increase, causing the observed spectrum to broaden and the per-bin power to be redistributed over a wider angular region even for the same underlying channel realization. In the isotropic case, the spectrum approaches a flat response reflecting isotropic reception on the scan grid. These examples visualize why simple power summation over scan angles is biased without compensating for the overlap effect, motivating the correction in \eqref{eq:Pc_hat}.

To evaluate the accuracy of the proposed method in simulation, we form the multi-dimensional power vector by taking the element-wise squared magnitude of the synthesized MDCIR samples and then
compare the estimated synthesized channel power with the true channel power obtained from the generated multipath components. Here, for each HPBW setting, we generated 1,000 independent channel realizations. For each realization, we performed 100 trials by applying i.i.d.\ random phases to all MPCs to enforce the US assumption, and the reported results are averaged over trials. The synthesized omnidirectional channel power is then estimated by summing over the scan grid and compensating for beam overlap using the beam-accumulation factor as
\begin{equation}
\widehat{P}_c = \frac{1}{\zeta}\sum_{k=1}^{N_{\tot}} [\Vect{P}]_k,
\label{eq:Pc_hat}
\end{equation}
where $\zeta$ is computed on the same discrete scan grid using the adopted Gaussian beam model. We evaluate two compensation variants: (i) the on-grid (bin-centered) accumulation factor $\zeta$, and (ii) the within-bin averaged factor $\bar{\zeta}$ that accounts for off-grid directions (scalloping). For each HPBW value, the estimation error is evaluated in dB as
\begin{equation}
\varepsilon_{\rm dB} \triangleq 10\log_{10}\!\left(\frac{\mathbb{E}[ \widehat{P}_c]}{P_c}\right),
\end{equation}
where \revision{yellow}{(R1-1)}{}{$P_c$ is defined in \eqref{eq:US}, and $\mathbb{E}[\cdot]$ denotes averaging over the random-phase Monte-Carlo trials. It should be emphasized that $P_c$ serves as an oracle reference because it is calculated directly from the true powers of the MPCs used to generate each channel realization, i.e., $P_c=\sum_{l=1}^{L}|\gamma_l|^2$. Thus, no MPC parameter estimation is involved in obtaining the reference channel power.

This oracle reference corresponds to the ideal power estimate that a high-resolution MPC parameter-estimation method such as SAGE or CLEAN would obtain if all MPCs were perfectly identified and their powers were perfectly estimated. In practice, the power reconstructed by such methods may additionally depend on the adopted model order, initialization, stopping criterion, detection threshold, and the treatment of unresolved or diffuse multipath components. Accordingly, comparison with the exact MPC power sum $P_c$ provides a direct benchmark for evaluating the accuracy of the proposed synthesized-power recovery without introducing implementation-dependent MPC parameter-estimation errors.}

Fig.~\ref{fig:estimation_error} summarizes the Monte-Carlo results of $\varepsilon_{\rm dB}$ for several directional scanning configurations. The reference curve computed from the true MPC powers remains close to $0$~dB, validating the Monte-Carlo averaging under the uncorrelated scattering assumption. The correction using the on-grid accumulation factor $\zeta$ mitigates the dominant overlap bias but can leave a residual offset when MPC directions are off-grid (scalloping) and/or when the Co-El scan grid suffers from pole-related discretization artifacts. Using the averaged factor $\bar{\zeta}$ further reduces this residual bias, improving robustness with respect to HPBW and scan coarseness, particularly in wide-beam cases and in Co-El scanning.

\subsection{Measurement Based Evaluation}
\label{subsec:meas_evaluation}
\begin{figure}[t]
\centering
\includegraphics[width=\linewidth]{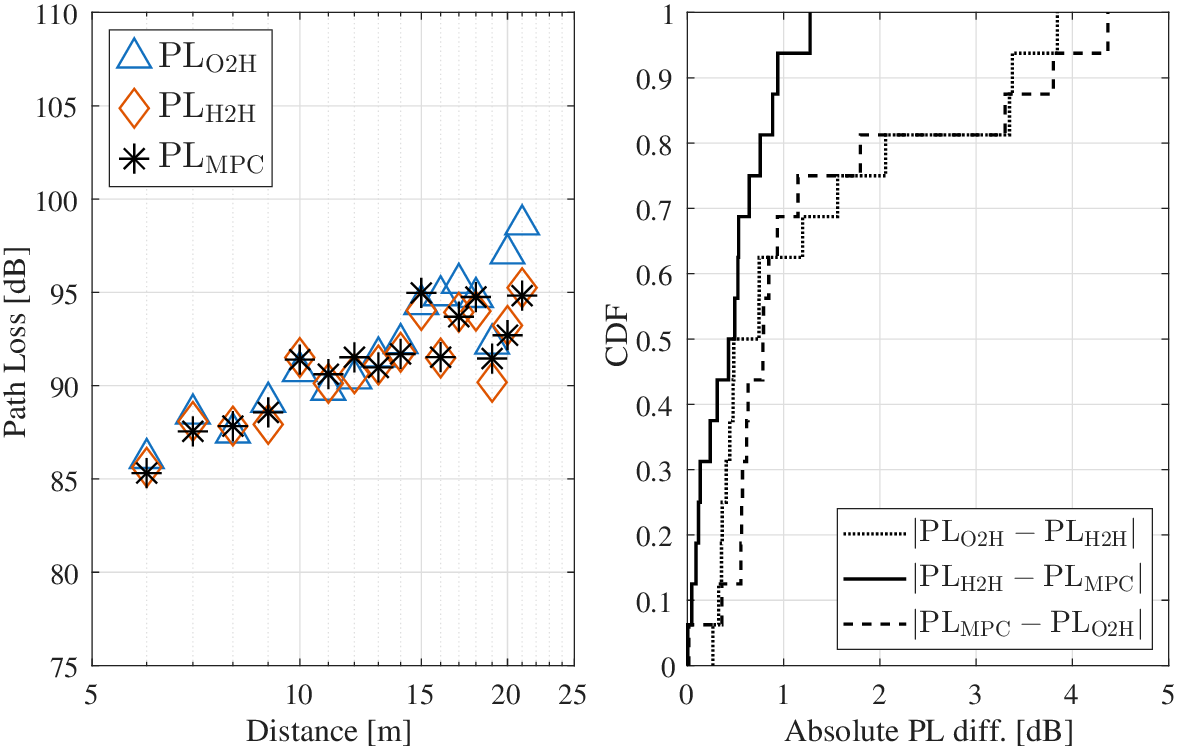}
\caption{\revision{yellow}{(R1-1)}{}{Comparison of three independently obtained PL estimates: (a) PL versus Tx--Rx separation for the O2H measurement, the proposed beam-accumulation-corrected H2H synthesis, and the CLEAN-based MPC power summation; and (b) CDFs of the absolute pairwise PL differences among the three estimates.}}
\label{fig:PL_eval}
\end{figure}

\revision{yellow}{(R1-1)}{}
{To further validate the proposed method using measured data, we compare three independently obtained PL estimates: (i) the PL obtained from the horn-to-omnidirectional (O2H) measurement, in which only the Rx horn is angle-scanned and the proposed beam-accumulation correction is applied; (ii) the PL synthesized from the D-D horn-to-horn (H2H) measurement, in which both the Tx and Rx horns are angle-scanned and the proposed beam-accumulation correction is applied; and (iii) the PL calculated from the incoherent sum of the MPC powers extracted from the H2H measurement using the CLEAN algorithm \cite{Kim_CLEAN}. Unlike the simulation-based validation in Section~\ref{sec:simbasedeval}, the true MPC parameters and hence the exact propagation-channel power are not available in the measurement. Therefore, none of the three estimates is treated as an exact ground-truth reference; instead, their mutual consistency is evaluated.

Measurements were conducted in a line-of-sight (LoS) corridor at 154~GHz \cite{Corridor300GHz}, with the Tx--Rx separation swept from $6$--$21$~m in 1-m steps. For the H2H measurement, both ends used 26-dBi horns (HPBW $\approx 9^\circ$) shown in Fig.~\ref{fig:Horn_patterns} and were rotated over $360^\circ$ with a $9^\circ$ angular sampling interval. For the O2H measurement, the Tx horn was replaced with a 3-dBi omnidirectional antenna (Eravant, SAO-1141740330-06-S1) \cite{Eravant_Omni}, while only the Rx horn was rotated over $360^\circ$ with the same $9^\circ$ angular sampling interval.

Based on \eqref{eq:Pc_from_zeta}, the PLs obtained from the proposed synthesis are calculated as
\begin{align}
\mathrm{PL}_\mathrm{O2H} &= -10\log_{10}\!\Big(\frac{1}{\zeta_{\phi_\R}}\sum_{\check{\tau},\check{\phi}_\R} P(\check{\tau},\check{\phi}_\R)\Big),\\
\mathrm{PL}_\mathrm{H2H} &= -10\log_{10}\!\Big(\frac{1}{\zeta_{\phi_\T}\zeta_{\phi_\R}} \sum_{\check{\tau},\check{\phi}_\T,\check{\phi}_\R} P(\check{\tau},\check{\phi}_\T,\check{\phi}_\R)\Big),
\label{eq:PL_sum_meas}
\end{align}
where the within-bin averaged correction factors are employed. For comparison with a conventional high-resolution parameter-estimation approach, the CLEAN algorithm is also applied to the measured H2H angle-resolved channel responses. The number of MPCs extracted by CLEAN ranged from $3$ to $13$. Let $\widehat{p}_l$ denote the estimated power of the $l$th MPC extracted by CLEAN. The corresponding CLEAN-based PL is calculated from the incoherent sum of the extracted MPC powers as
\begin{equation}
\mathrm{PL}_\mathrm{MPC} = -10\log_{10}
\left(
\sum_{l=1}^{\widehat{L}}\widehat{p}_l\right),
\label{eq:PL_sum_meas}
\end{equation}
where $\widehat{L}$ denotes the number of extracted MPCs. Ideally, the three estimates ${\rm PL}_{\rm O2H}$, ${\rm PL}_{\rm H2H}$, and ${\rm PL}_{\rm MPC}$ should coincide if the beam-overlap compensation, the O2H measurement, and the CLEAN-based MPC extraction are all ideal. Fig.~\ref{fig:PL_eval}(a) compares the three PL estimates as a function of the Tx--Rx separation. The three results exhibit generally consistent distance-dependent trends over most of the measured range.

To assess their mutual consistency without selecting any one estimate as the ground-truth reference, Fig.~\ref{fig:PL_eval}(b) presents the CDFs of the absolute pairwise PL differences, $|{\rm PL}_{\rm O2H}-{\rm PL}_{\rm H2H}|,\quad |{\rm PL}_{\rm H2H}-{\rm PL}_{\rm MPC}|,\quad |{\rm PL}_{\rm MPC}-{\rm PL}_{\rm O2H}|$. At the longest Tx--Rx separations, the H2H result and the CLEAN-based result remain relatively close, whereas the O2H result tends to deviate more noticeably from both. This behavior is consistent with increasing SNR degradation in the O2H measurement. In the O2H configuration, the 26-dBi Tx horn used in the H2H measurement is replaced by a 3-dBi omnidirectional antenna, resulting in an approximately 23-dB lower Tx-side antenna gain. Consequently, the O2H measurement becomes increasingly susceptible to the noise floor as the separation distance increases.

The agreement between the proposed H2H synthesis and the CLEAN-based result provides an additional measurement-based validation of the proposed power-recovery method. At the same time, unlike CLEAN, the proposed method obtains the synthesized-omnidirectional channel power directly from the measured multidimensional power response without requiring explicit estimation of individual MPC parameters.
}

\section{Conclusion}
This paper investigated antenna-independent channel parameter extraction (e.g., received power and PL) from angle-resolved wideband channel measurements, focusing on angular-domain bias in practical angle-scanning. While delay-domain power integration is largely straightforward after back-to-back calibration, naive angular summation is biased due to non-orthogonal, overlapping scan beams, and is further affected by scalloping for off-grid angles under a finite ASI. To address this, we formulated the synthesized-isotropic narrowband received power in a matrix form and introduced a beam-accumulation correction factor computed via discrete scan-grid summation of the squared antenna pattern, together with an offset-averaged variant to reduce scalloping sensitivity without explicit off-grid angle estimation. \revision{yellow}{(R1-1)}{}{Simulations verified reduced overlap-induced bias and improved robustness versus HPBW and ASI, while 154~GHz LoS corridor measurements showed consistent PL trends among the proposed H2H synthesis, the O2H measurement, and the CLEAN-based MPC power summation.}

\balance


\begin{thebibliography}{00}
\bibitem{Rappaport}
T. S. Rappaport, \textit{Wireless Communications: Principles and Practice}, 2nd ed. Upper Saddle River, NJ, USA: Prentice Hall, 2002.

\bibitem{Erceg} 
V. Erceg, L. J. Greenstein, S. Y. Tjandra, S. R. Parkoff, A. Gupta, B. Kulic, A. A. Julius, and R. Bianchi, ``An Empirically Based Path Loss Model for Wireless Channels in Suburban Environments,'' \textit{IEEE J. Sel. Areas Commun.}, vol.~17, no.~7, pp.~1205--1211, Jul. 1999.

\bibitem{ITU_R_P_341}
ITU-R, ``The concept of transmission loss for radio links,'' Recommendation ITU-R P.341-7, Aug. 2019.

\bibitem{ITU_R_P_1407}
ITU-R, ``Multipath propagation and parameterization of its characteristics,'' Recommendation ITU-R P.1407-8, Sept. 2021.

\bibitem{Steinbauer}
M. Steinbauer, A. F. Molisch, and E. Bonek, ``The double-directional radio channel,'' \textit{IEEE Antennas Propag. Mag.}, vol.~43, no.~4, pp.~51--63, Aug. 2001.

\bibitem{KIM_11GIndoor} 
M. Kim, Y. Konishi, Y. Chang, and J. Takada, ``Large Scale Parameters and Double-Directional Characterization of Indoor Wideband Radio Multipath Channels at 11 GHz," \textit{IEEE Trans. Antennas Propag.}, vol.~62, no.~1, pp.~430--441, Jan. 2014.

\bibitem{Zwick}
T. Zwick, D. Hampicke, A. Richter, G. Sommerkorn, R. Thom\"{a}, W. Wiesbeck, ``A Novel Antenna Concept for Double-Directional Channel Measurements,'' \textit{IEEE Trans. Veh. Technol.}, vol.~53, no.~2, pp.~527--537, Mar. 2004.

\bibitem{Kim_IEEEAccess2021_FullAzSweep}
M. Kim, S. Tang, and K. Kumakura,
``Fast Double-Directional Full Azimuth Sweep Channel Sounder Using Low-Cost COTS Beamforming RF Transceivers,'' \textit{IEEE Access}, vol.~9, pp.~80288--80299, 2021.

\bibitem{Thoma}
R. Thom\"{a}, M. Landmann, A. Richter, and U. Trautwein, ``Multidimensional High-Resolution Channel Sounding,'' in Smart Antennas in Europe State of the Art, EURASIP Book Series on SP \& C, K. T., Ed. Hindawi Publishing Corporation, 2005, vol. 3.

\bibitem{Doeker_OJAP2024_Calibration}
T. Doeker, J. M. Eckhardt, C. E. Reinhardt, and T. K\"urner, ``Time-Domain Channel Sounder Calibration at Low Terahertz Band,'' \textit{IEEE Open J. Antennas Propag.}, vol.~5, no.~6, pp.~1598--1611, 2024.

\bibitem{Fleury_SAGE}
B. Fleury, M. Tschudin, R. Heddergott, D. Dahlhaus, and K. Pedersen, ``Channel Parameter Estimation in Mobile Radio Environments Using the SAGE Algorithm,'' \textit{IEEE J. Sel. Areas Commun.}, vol.~17, no.~3, pp.~434--450, 1999.

\bibitem{Richter_RIMAX}
A. Richter, ``Estimation of Radio Channel Parameters: Models and Algorithms,'' Ph.D. dissertation, TU Ilmenau, 2005.

\bibitem{Kim_CLEAN}
M. Kim, T. Iwata, S. Sasaki, J. Takada, ``Millimeter-Wave Radio Channel Characterization using Multi-Dimensional Sub-Grid CLEAN Algorithm,'' \textit{IEICE Trans. Commun.}, vol. E103-B, no.~7, pp.~767--779, Feb. 2020.

\bibitem{Kim_Springer}
M. Kim, ``Millimeter-wave propagation of 5G wireless access,'' in \textit{Handbook of Radio and Optical Networks Convergence}, T. Kawanishi, Ed. Singapore: Springer, 2023, pp.~1--28.

\bibitem{Kaske}
M. K\"{a}ske, M. Landmann, and R. Thom\"{a}, ``Modelling and Synthesis of Dense Multipath Propagation Components in the Angular Domain,'' \textit{the 3rd European Conference on Antennas and Propagation (EuCAP)}, pp.~2641--2645, Berlin, Germany, Mar. 2009. 

\bibitem{Sun_Globcom2015_OmniSynth}
S. Sun, G. R. MacCartney~Jr., M. K. Samimi, and T. S. Rappaport, ``Synthesizing Omnidirectional Antenna Patterns, Received Power and Path Loss from Directional Antennas for 5G Millimeter-Wave Communications,''
in \textit{Proc. IEEE GLOBECOM}, Dec. 2015.

\bibitem{Haneda_EuCAP2016_OmniPL}
K. Haneda, S. L. H. Nguyen, J. J\"arvel\"ainen, and J. Putkonen, ``Estimating the omni-directional pathloss from directional channel sounding,'' in \textit{Proc. 10th Eur. Conf. Antennas Propag. (EuCAP)}, 2016.

\bibitem{Daoud}
D. Burghal, J. Gomez-Ponce, N. A. Abbasi and A. F. Molisch, ``Estimation of Isotropic Pathloss From Directional Channel Measurements in Azimuth and Elevation,'' \textit{IEEE Antennas Wirel. Propag. Lett.}, vol.~23, no.~7, pp.~2145--2149, Jul. 2024.


\bibitem{Xu}
H. Xu, J. Zhang, P. Tang, H. Xing, L. Tian and Q. Wang, ``High-Resolution Multipath Angle Estimation Based on Power-Angle-Delay Profile for Directional Scanning Sounding,'' \textit{IEEE Trans. Antennas Propag.} (early access).

\bibitem{Li_Fan_TVT2023}
M. Li, F. Zhang, X. Zhang, Y. Lyu and W. Fan, ``Omni-Directional Pathloss Measurement Based on Virtual Antenna Array With Directional Antennas,'' \textit{IEEE Trans. Veh. Tech.}, vol.~72, no.~2, pp.~2576--2580, Feb. 2023.

\bibitem{Bello} 
P. A. Bello, ``Characterization of Randomly Time-variant Linear Channels,'' \textit{IEEE Trans. Commun. Syst.}, vol.~56, no.~12, pp.~360--393, Dec. 1963.

\bibitem{Kim_AWPL}
M. Kim, ``Analysis of Multipath Component Parameter Estimation Accuracy in Directional Scanning Measurement,'' \textit{IEEE Antennas Wireless Propag. Lett.}, vol.~17, no.~1, pp.~12--16, Jan. 2018.

\bibitem{Corridor300GHz}
R. Takahashi, A. Ghosh, M. Mao, M. Kim, ``Channel Modeling and Characterization of Access, D2D and Backhaul Links in a Corridor Environment at 300 GHz,'' \textit{IEEE Trans. Antennas Propag.}, vol.~73, no.~4, pp.~1954--1968, Apr. 2025.

\bibitem{NU_SC}
M. Kim, M. Yomoda, M. Mao, N. Kuno, K. Kitao, and S. Suyama, ``Quasi-Deterministic modeling of Sub-THz band access channels in street canyon environments,'' \textit{IEEE Trans. Antennas Propag.}, vol.~74, no.~7, pp.~7007--7021, Jul. 2026.

\bibitem{SV}
A. Saleh and R. Valenzuela, ``A Statistical Model for Indoor Multipath Propagation,'' \textit{IEEE J. Sel. Areas Commun.}, vol.~5, no.~2, pp.~128--137, 1987.

\bibitem{Eravant_Omni}
Eravant, ``Omnidirectional Antennas,'' [Online]. Available:
\url{https://www.eravant.com/products/antennas/omnidirectional-antennas}

\end{thebibliography}
\end{document}